\documentclass[lettersize,journal]{IEEEtran}
\usepackage{amsmath,amsfonts}
\usepackage{amssymb}
\usepackage{algorithmic}
\usepackage{algorithm}
\usepackage{array}
\usepackage{textcomp}
\usepackage{stfloats}
\usepackage{url}
\usepackage{verbatim}
\usepackage{graphicx}
\usepackage{cite}
\usepackage{booktabs}
\usepackage{multirow}
\usepackage{threeparttable}
\usepackage{makecell}
\usepackage{wasysym}
\usepackage{balance}
\usepackage{xcolor}
\usepackage{subcaption}
\begin{document}

\title{\textcolor{black}{Learning Time-Graph Frequency Representation \\ for Monaural Speech Enhancement}\\
\author{Tingting Wang, Tianrui Wang, Meng Ge, Qiquan Zhang, \IEEEmembership{Member, IEEE}, and 
Xi Shao, \IEEEmembership{Member, IEEE}
\thanks{This work was supported by the research and development of key technologies of secure, reliable and high-performance speaker identity verification platform for complex scenario, Jiangsu (Grant BG2024027). \textit{(Tingting Wang and Qiquan Zhang contributed equally to this work.) (Corresponding author: Qiquan Zhang.)}}
\thanks{Tingting Wang and Xi Shao are with the School of Communication and Information Engineering, Nanjing University of Posts and Telecommunications, Nanjing, China (e-mail:\{tingting\_wang, shaoxi\}@njupt.edu.cn).}
\thanks{Tianrui Wang and Meng Ge are with the Tianjin Key Laboratory of Cognitive Computing and Application, College of Intelligence and Computing, Tianjin University, Tianjin, China (e-mail: \{wangtianrui, gemeng\}@tju.edu.cn).}
\thanks{Qiquan Zhang is with the School of Electrical Engineering and Telecommunications, The University of New South Wales, Sydney, NSW 2052, Australia (e-mail: {zhang.qiquan}@outlook.com).}}}




\maketitle

\begin{abstract}
\textcolor{black}{The Graph Fourier Transform (GFT) has recently demonstrated promising results in speech enhancement. However, existing GFT-based speech enhancement approaches often employ fixed graph topologies to build the graph Fourier basis, whose the representation lacks the adaptively and flexibility. In addition, they suffer from the numerical errors and instability introduced by matrix inversion in GFT based on both Singular Value Decomposition (GFT-SVD) and Eigen Vector Decomposition (GFT-EVD). Motivated by these limitations, this paper propose a simple yet effective learnable GFT-SVD framework for speech enhancement. Specifically, we leverage graph shift operators to construct a learnable graph topology and define a learnable graph Fourier basis by the singular value matrices using 1-D convolution (Conv-1D) neural layer. This eliminates the need for matrix inversion, thereby avoiding the associated numerical errors and stability problem. In contrast to complex-valued representation, our proposed learnable Fourier basis provides a real-valued time-graph representation, enabling better magnitude–phase alignment in speech enhancement. Comprehensive evaluations on the VCTK+DEMAND and DNS-2020 benchmarks demonstrate the consistent performance superiority of our learnable GFT-SVD over fixed STFT, GFT-EVD, and GFT-SVD within existing neural speech enhancement frameworks. Source code is available at \url{https://github.com/Wangfighting0015/GFT\_project}.}


 
\end{abstract}


\section{Introduction}
\IEEEPARstart{I}{n} \textcolor{black}{real-world acoustic environments, a wide range of speech applications, such as hearing aids, automatic speech recognition (ASR), speaker verification, and brain-computer interfaces (BCIs), are inevitably affected by background noises. To mitigate this issue, speech enhancement (SE) has attracted substantial research interest, seeking to isolate the clean speech from a noisy mixture to improve the perceived speech quality and intelligibility~\cite{loizou,TimeFrequency,R1,R3,R4, R12,mambadc}. Traditional unsupervised SE methods mainly include spectral subtraction, Wiener filtering, and statistical model-based approaches~\cite{mmse2017,mmse,zhang2019,zhang2019fast}, which fail to suppress highly non-stationary noise sources. Over the past decade, the advent of deep neural networks (DNNs) has facilitated the remarkable success of supervised SE, demonstrating substantial superiority over traditional approaches~\cite{overview2018,SpeechEnhancement,DeepMMSE,chen2023neural,mamba}.}



\textcolor{black}{The initial neural SE typically leverages DNNs to estimate the clean spectral magnitude~\cite{XuDDL15,R5,R9} or a real-valued spectral mask~\cite{WangNW14,R6,R7,Performance}, given a noisy spectral magnitude. The estimated spectral mask is applied to the noisy spectral magnitude to obtain the clean spectral magnitude. The waveform of clean speech is then reconstructed using the noisy phase. Later, researchers have demonstrated the importance of phase modeling for speech intelligibility~\cite{PaliwalWS11}, leading to the emergence of neural phase-aware speech enhancement~\cite{8466892,R10,R11}.} These methods can be broadly grouped into waveform-based~\cite{SEGAN,Endtoend,Onlossfunctions,TCNN,LuoM19} and spectrogram-based schemes~\cite{psm,CRM,PHASEN}. The former separates the clean waveform directly from the noisy waveform in an end-to-end manner, with phase modeling implicitly involved in waveform alignment learning. In contrast, the latter explicitly models the phase information via either spectral masking or mapping approaches. The complex spectral masking methods typically predict a complex-valued ratio mask (CRM)~\cite{CRM} or jointly predict a spectral magnitude mask and a phase mask (or phase spectrum)~\cite{MPSENet, PHASEN} for complex noisy spectrum, which simultaneously reconstruct magnitude and phase. The complex spectral mapping optimizes DNNs to directly predict the real and imaginary parts of clean speech spectrum~\cite{csm,bsrnn,DCCRN}. 
\textcolor{black}{However, complex spectral mapping and masking approaches inevitably sacrifice magnitude estimation accuracy to compensate for phase prediction fidelity, particularly under low signal-to-noise ratios (SNR)~\cite{WangWR21}.}

\textcolor{black}{To mitigate this issue, a decoupling-style spectral mapping approach has been proposed~\cite{Two-Stage}, which disentangles complex spectrum estimation into two stages: magnitude reconstruction followed by the residual complex spectrum estimation. This allows for separate optimization spaces for modeling magnitude and phase, alleviating the magnitude-phase compromise in joint estimation paradigms~\cite{LiYYZL22}. \textcolor{black}{Subsequently, the study~\cite{DBTNet} proposes a dual-branch collaborative framework to recover the magnitude and phase in parallel, with cross-branch information interaction leveraging spectral magnitude-complexity associations. }Nevertheless, this decoupling process raises a critical challenge—the effective alignment between magnitude and phase modeling in STFT-based neural SE.}

\textcolor{black}{In contrast to the STFT, which uses a fixed Fourier basis, the Graph Fourier Transform (GFT) flexibly constructs the time-graph representation via adjacency matrices, enabling task-specific representation. The STFT imposes a fixed frequency resolution across the spectrum, whereas human auditory system is inherently non-uniform, the GFT provides the flexibility with configurable graph topology to better align with perceptual frequency scales. \textcolor{black}{Recent studies leveraging the GFT for audio and speech tasks, including audio watermarking~\cite{ref133}, speech anti-spoofing~\cite{xu22_odyssey}, and SE~\cite{YanYWG20,ZhangP22,timegraph}, have demonstrated promising results.} Yan~\textit{et al.}~\cite{YanYWG20} propose a GFT based on Eigenvalue Decomposition (GFT-EVD) that maps a noisy waveform into a complex-valued time–graph representation for magnitude spectral subtraction, achieving better results than the STFT counterpart. The graph Laplacian matrix enables the model to capture the intrinsic relationships among speech frames more effectively than the STFT~\cite{ZhangP22}. Wang \textit{et al.}~\cite{timegraph} propose a GFT based on Singular Value Decomposition (GFT-SVD) that maps a noisy waveform into a real-valued time-graph representation, eliminating the need for explicit amplitude-phase alignment.}

\textcolor{black}{However, existing GFT-based SE approaches often rely on a fixed graph topologies, which limits their adaptability to the dynamic speech structures~\cite{newmultilayer,WangGGZY23}. Moreover, they suffer from numerical errors and instability problem introduced by matrix inversion involved in calculation of GFT. In this paper, we propose a learnable GFT-SVD to enable joint optimization of the time-graph representation and the mask learning backbone networks in an end-to-end (E2E) fashion. Specifically, we exploit the graph shift operator to build a learnable graph topology for the learnable graph Fourier basis. Furthermore, we propose using 1-D convolution neural layer to learn the inverse process of matrix. In summary, the main contributions of this paper are as follows:}

\begin{itemize}
  \item \textcolor{black}{We propose a simple yet effective learnable GFT-SVD representation for E2E speech enhancement. It can be easily integrated with existing speech enhancement backbones to further improve performance.}
  \item \textcolor{black}{We design a learnable graph topology with a learnable Fourier basis, jointly optimized with mask learning to provide an adaptive speech representation. This formulation eliminates the reliance on matrix inversion, thereby avoiding numerical errors and stability issues. Furthermore, the resulted real-valued time-graph representation enables better magnitude-phase alignment than complex-valued representation in speech enhancement.}
  \item \textcolor{black}{We extensively evaluate the learnable GFT-SVD across diverse backbone networks, on two widely used benchmarks, i.e., VCTK+DEMAND and DNS-2020. The experimental results consistently confirm the improvements in speech intelligibility and perceptual quality.}
\end{itemize}

\textcolor{black}{The remainder of this paper is structured as follows. Section~II presents and discuss the preliminary work. Section III details our proposed learnable graph topology and learnable GFT-SVD for speech enhancement. Section IV describes the experimental setup. The experimental results are presented and discussed in Section V. Finally, Section VI concludes this paper.} 

\section{Preliminary Work}\label{Related_Works}

\subsection{\textcolor{black}{Graph Signal Model}}
\textcolor{black}{Given a clean speech signal $\boldsymbol{s}$ degraded by an uncorrelated noise signal $\boldsymbol{d}$, the observed noisy speech $\boldsymbol{x}$ can be formulated as:
\begin{equation}
  \boldsymbol{x}[n]=\boldsymbol{s}[n]+\boldsymbol{d}[n]
  \label{sec:typestyle},
\end{equation}
where $n$ denotes the index of discrete-time samples. The noisy speech signal $\boldsymbol{x}$ is then represented as a graph speech signal $\boldsymbol{x}_{\cal{G}}$ via vertex-wise assignment, where the value $\boldsymbol{x}_{\cal{G}}[v_{n}]$ at the $n$-th vertex $v_n$ corresponds to the $n$-th sample $\boldsymbol{x}[n]$, thereby 
establishing a injection mapping:
 \begin{equation}
    x:{\mathbb{R}} \to {{\cal V}},{x[n]} \to {\boldsymbol{x}_{\cal{G}}[v_n]}
\end{equation}
where ${\cal V}$ denotes the vertex set. The graph topology ${\cal{G}}_x$ of $\boldsymbol{\bf x}_{\cal{G}}\in {\mathbb{R}^{N}}$ is formally defined as:
\begin{equation}
   {{\cal{G}}_x} = \left( {\cal V}, {\boldsymbol {\cal E}_K}, {\bf{W}}\right),
   \label{sec:typestyle}
\end{equation}
where $\boldsymbol{{\cal E}}_{K}$ denotes the set of $K$ edges connecting each speech sample to its $K$ surrounding samples, which determines the initial sparsity level of the graph topology. The number of vertexes is $ N = |{\cal V}|$ denoting the length of speech frame~\cite{ref8,ref9}. The $k$-th edge ${{{\cal{E}}}_k}(n, m)$ is set to 1 if there exists a dependency between the $n$-th graph speech sample $\boldsymbol{x}_{\cal{G}}[v_n]$ and the $m$-th graph speech sample $\boldsymbol{x}_{\cal{G}}[v_m]$, and $0$ otherwise. ${\bf W}\!\in\!{\mathbb{R}^{N \times N}}$ denotes the graph shift operator, with the value of ${\bf{W}}(n, m)$ indicating the strength of dependency between $\boldsymbol{x}_{\cal{G}}[v_{n}]$ and $\boldsymbol{x}_{\cal{G}}[v_{m}]$~\cite{ref1, ref9}. The graph topology is initialized according to the edge set $\boldsymbol{\mathcal{E}}_K$. For instance, with $K\!=\!3$, ${\bf W}$ takes the following form:}
\begin{equation}
{{\bf W}}{\rm{ = }}\left[ {\begin{array}{*{20}{c}}
0&0&0&0& \cdots&1&1&1\\
1&0&0&0& \cdots &0&1&1\\
1&1&0&0& \cdots &0&0&1\\
 \vdots & \vdots & \vdots & \vdots & \cdots & \vdots & \vdots & \vdots \\
0&0&0&0& \cdots & \cdots&1&0
\end{array}} \right].
\end{equation}

\subsection{\textcolor{black}{Speech Spectrum Transform}}
\begin{figure*}[t]
\centering 
\includegraphics[width=1.0\linewidth]{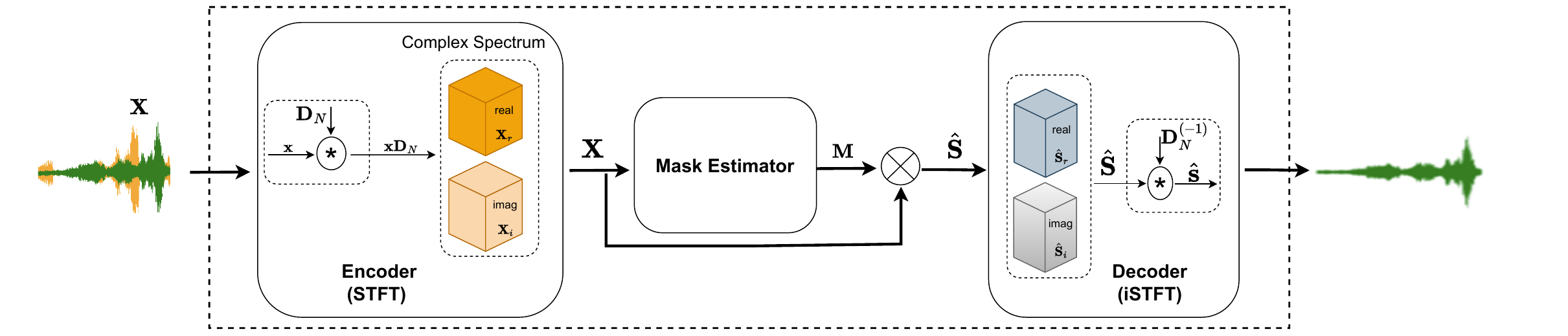}
 \caption{\textcolor{black}{The STFT-based neural speech enhancement architecture, where ${\bf D}_N$ represents the classical Fourier matrix.}}
	\label{framework1}
\end{figure*}

\textcolor{black}{\textbf{STFT.} The noisy speech waveform $\boldsymbol{x}$ can be transformed into a complex-valued time-frequency spectrogram via the short-time Fourier Transform (STFT):
\begin{equation}
{{{{\textbf X}}}}=\text{STFT}{({\boldsymbol x})} = {\textbf D}_N{\boldsymbol x}= {{\textbf X}_r + j{\textbf X}_i}{\in \mathbb{C}} 
\end{equation}
where ${\textbf D}_N$ is the classical Fourier matrix. ${{\textbf X}_r}$ and ${{\textbf X}_i}$ denote the real and imaginary parts of the complex-valued spectrogram, respectively. \textcolor{black}{A typical neural approach to STFT-based speech enhancement optimizes a DNN with two dedicated prediction heads that respectively estimate the real and imaginary parts of clean speech spectrum, illustrated as in Fig.~\ref{framework1}.}}

\textcolor{black}{\textbf{GFT-EVD.} Given the graph speech signal $\boldsymbol{x}_{\cal{G}}$, its Graph Fourier Transform (GFT) is obtained via the eigenvalue decomposition (EVD) of the graph adjacency matrix $\mathbf{A}$, formulated as:
\begin{equation}
\mathbf{X}_{\cal{G}} = \text{GFT-EVD}(\boldsymbol{x}_{\cal{G}}) = \mathbf{U}^{-1} \boldsymbol{x}_{\cal{G}} = \mathbf{X}_{\cal{G}}^r + j \mathbf{X}_{\cal{G}}^i \in \mathbb{C},
\end{equation}
\begin{equation}
\rm{EVD}\left(\mathbf{A}\right) = \mathbf{U} \boldsymbol{\Lambda} \mathbf{U}^{T}
\end{equation}
where $\mathbf{U}$ denotes the eigenvector matrix of $\mathbf{A}$, and the diagonal elements of the diagonal matrix $\boldsymbol{\Lambda} = \text{diag}(\lambda_0, \lambda_1, \ldots, \lambda_{N-1})$ denote the graph frequencies. The superscript $T$ denotes the conjugate transpose of a matrix, and ${\mathbf{X}}_{\cal{G}}^r$ and $\mathbf{X}_{\cal{G}}^i$ denote the real and imaginary components of the GFT-EVD. Similar to STFT, GFT-EVD transforms the graph speech signal into a complex-valued spectrum, and the inverse GFT-EVD reconstructs the graph speech via the eigenvector matrix $\mathbf{U}$:
\begin{equation}
{{{{\boldsymbol x_{\cal{G}}}}}} = {\text{iGFT-EVD}{({{\textbf X}_{\cal{G}}}})} = {\bf U}{{\textbf X}_{\cal{G}}}
\end{equation}}

\textcolor{black}{\textbf{GFT-SVD}. Unlike the STFT and GFT-EVD, the singular value decomposition (SVD) based GFT (GFT-SVD) transforms the graph speech signal $x_{\cal G}$ into a real-valued spectrum representation $\textbf X_{\cal{G}}$, given by:
\begin{equation}
{{{{\textbf X_{\cal{G}}}}}} = \text{GFT-SVD}{({\boldsymbol x}_{\cal G })}={\bf \Psi^{-1}}{\boldsymbol x_{\cal G}} \in {\mathbb R} 
\end{equation}
\begin{equation}
{\rm{SVD}}\left({{\bf A}} \right) = {\bf \Psi} \times {\bf \Lambda} \times {{\bf \Gamma}}{} 
\end{equation}
where ${\mathbf \Psi}$ denotes the left singular-vector matrix along the time dimension, which aligns with the phonetic boundaries of speech. ${\mathbf \Gamma}$ denotes the right singular-vector matrix that characterizes harmonics in the frequency domain. The inverse GFT-SVD reconstructs the graph speech signal via the left singular-vector matrix ${\bf \Psi}$:
\begin{equation}
{\boldsymbol {x}}_{\cal{G}} = \text{iGFT-SVD}({\bf {X}}_{{\cal{G}}}) ={\bf \Psi}{\textbf{X}}_{\cal{G}}.
\label{eq:decoder}
\end{equation}
In contrast to the STFT, which operates in a fixed Fourier basis, the GFT can be flexibly constructed via various graph adjacency matrices, enabling task-specific representations. Similar to the STFT, the GFT-EVD yields a complex-valued spectral representation of the graph speech signal, presenting a magnitude–phase alignment challenge~\cite{YanYWG20}. In contrast, the GFT-SVD produces a real-valued spectral representation, thereby avoiding this issue. Furthermore, we propose introducing learnable neural layers to model the inversion process, aimed at mitigating numerical errors and instability issues arising from the matrix inversion in GFT~\cite{ZhangYKCWWE22, timegraph}. The details of our proposal are presented in Section~III.}

\section{Methodology}
\label{sec:methodology}

\subsection{Overview} \label{RLM}
\begin{figure*}[t]
	\centering
\includegraphics[width=1.0\linewidth]{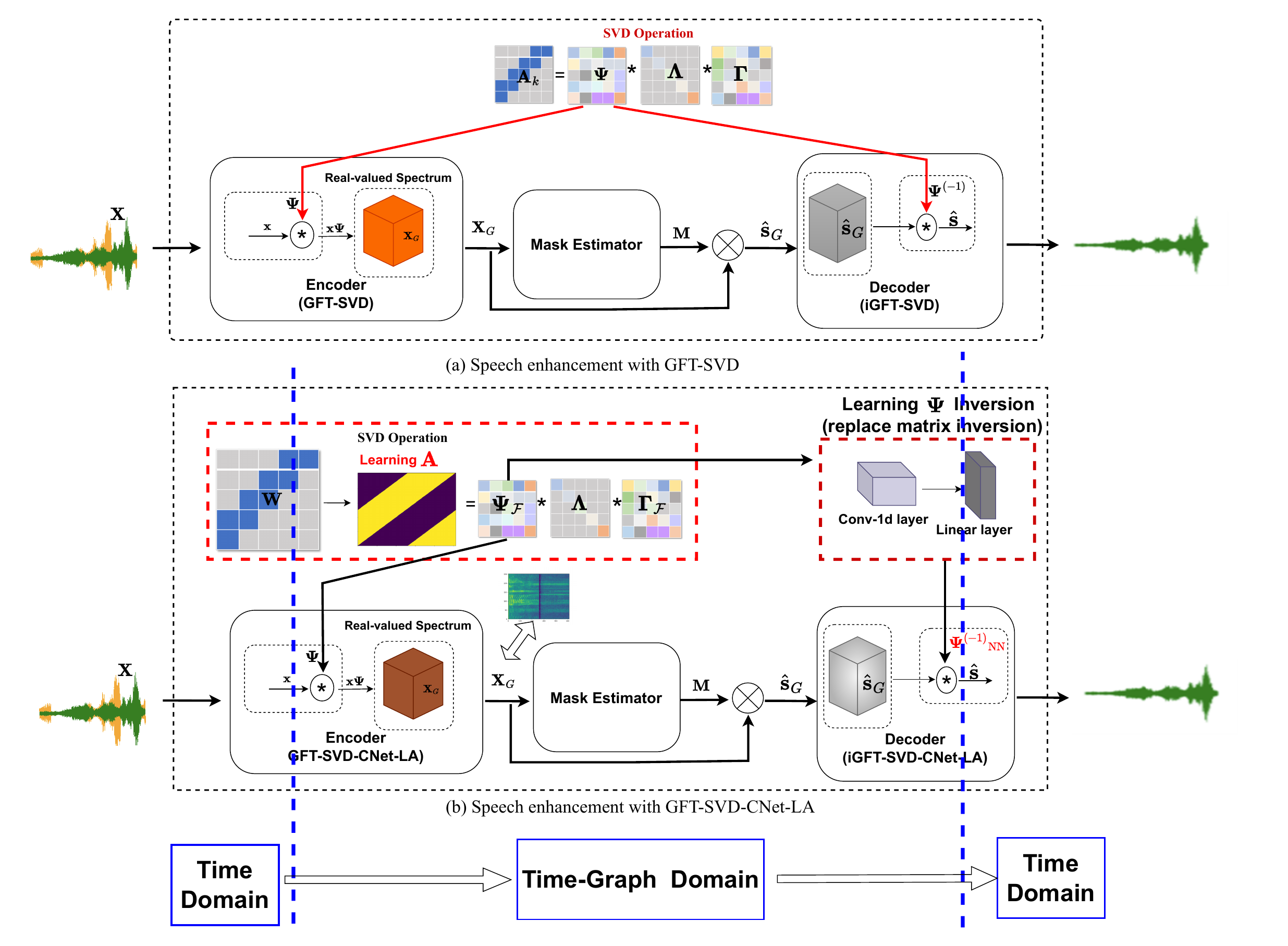}
	\caption{(a) The neural speech enhancement architecture with GFT-SVD. (b) \textcolor{black}{The overview of neural speech enhancement  architecture with learnable GFT-SVD.}}
	\label{framework}
	\vspace{-8pt}
\end{figure*}

\textcolor{black}{As illustrated in Fig.~\ref{framework} (b), the overall architecture of the proposed learnable GFT-based speech enhancement comprises a GFT-SVD encoder, a mask learning backbone, and a GFT-SVD decoder.} \textcolor{black}{Specifically, a graph shift operator is used to establish a learnable graph topology in the encoder module, enabling the graph adjacency matrix to dynamically characterize intrinsic relationships among graph speech samples. \textcolor{black}{Furthermore, we introduce the 1-D convolution (Conv-1D) layer for learnable inversion of graph Fourier basis in the decoder module, eliminating the need for explicit matrix inversion. In contrast to STFT-based frameworks, our proposal mitigates the compensation problem between magnitude and phase spectrum learning~\cite{WangWR21}, facilitating better magnitude–phase alignment for speech enhancement. The mask learning backbone can be flexibly implemented using existing enhancement networks.}}

\textcolor{black}{Formally, a learnable graph topology-based GFT-SVD encoder, referred to as GFT-SVD-LA, initialized according to the graph shift operator ${\bf W}$, maps a noisy speech signal $\boldsymbol{x} \in {\mathbb{R}}^{N}$ into a real-valued time–graph representation ${\bf X}_{\cal G} \in {\mathbb{R}}^{N}$. 
The representation ${\bf X}_{\cal G}$ is then passed to a mask learning backbone (mask estimator) that estimates the mask $\mathbf{M}$, which is applied to noisy speech graph representation, yielding ${\bf \hat{S}}_{\cal G} = {\bf M} \odot {\bf X}_{\cal G}$. The clean speech waveform is reconstructed from ${\bf \hat{S}}_{\cal G}$ through the inverse GFT-SVD-LA with Conv-1D, termed GFT-SVD-CNet-LA. The overall framework incorporating the learnable GFT-SVD is formulated as:
\begin{gather}
   {{{{\textbf X_{\cal{G}}}}}} = \text{GFT-SVD-LA}{({\boldsymbol x}_{\cal G })} \in \mathbb{R} \label{eq:encoder}\\
    {\bf \hat{S}}_{{\cal{G}}} ={\bf M} \odot {\bf X}_{\cal {G}}\\
    \hat{\textbf{s}} = \text{iGFT-SVD-CNet-LA}({\bf \hat{S}}_{{\cal{G}}}) \in \mathbb{R}. \label{eq:decoder}
\end{gather}}
The detailed workflow of the learnable graph topology and graph Fourier basis are described in the next subsection.
\subsection{The Learnable \textcolor{black}{ Graph Topology}}
\label{subsec:dvc} 
The speech samples \( \boldsymbol x \) are mapped to a graph signal \( \mathbf{  \boldsymbol x}_{\mathcal{G}} \) on a fixed graph topology \( \mathcal{G}_x = (\mathcal{V}, \mathcal{E}_K, \mathbf{W}) \). The adjacency matrix element \( \mathbf{A}(n,m) \) for each edge \( (n,m) \in \mathcal{E}_K \) is initialized by weights \( \mathbf{W} \), then adaptively weighted to dynamically model dependency strengths between \( \mathbf{x}_{\mathcal{G}}[v_n] \) and \( \mathbf{x}_{\mathcal{G}}[v_m] \). Assuming \( \mathbf{x}_{\mathcal{G}}[v_n] \) connects to its \( K \) nearest neighbors \( \mathbf{x}_{\mathcal{G}}[v_m] \), ${\mathbf {A}}(n,m)$ can be given as $\mathbf{A}(n,m)={\mathcal{F}}((x_{\cal G}[{v}_n], x_{\cal G}[{v}_m]), \quad \forall (n,m) \in \mathcal{E}_k)$ by employing a parameterized neural network \( \mathcal{F}(\cdot, \Theta) \). Each row of $\mathbf{A}$ is constrained to sum to unity, forming a row-stochastic matrix that represents valid probability distributions over connections. This property enables $\mathbf{A}$ to adaptively capture the intrinsic relationships among speech samples. The learnable graph topology is formulated as:
\begin{gather}
{{\cal{G}}_x} = \left( {\cal V}, \boldsymbol{{\cal E}}_{K}, {{\bf A}}\right)\\
    {\bf A} = {\mathcal{F}}(((x_{\cal G}[{v}_n], x_{\cal G}[{v}_m])| {\bf W}), \Theta ) \in \mathbb{R}^{ N \times N},\quad \forall (n,m) \in \mathcal{E}_K \\
    \sum_{m=0}^{\rm N-1}{{\rm A}(n,m)}=1,n \in {0,1, ..., {N-1}}. \label{eq:decoder1}
\end{gather}
\textcolor{black}{The learnable graph topology construction is shown in the black dashed area in the upper left of Fig. 2~(b).}



\subsection{\textcolor{black}{Learnable Real-Valued Graph Fourier Basis}}\label{dvc}
\label{subsec:dvc}

Given the learnable graph topology $\mathcal{G}_x$, we derive its adjacency matrix $\mathbf{A}$ and construct a learnable graph Fourier basis via Singular Value Decomposition (SVD), i.e., $\operatorname{SVD}(\mathbf{A}) = \boldsymbol{\Psi}_{\mathcal{F}} \boldsymbol{\Lambda} \boldsymbol{\Gamma}_{\mathcal{F}}$, where $\bf {\bf \Psi}_{\cal{F}}$ and $\bf {\bf \Gamma}_{\cal{F}}$ denote the left and right singular vector matrices of $\bf A$.
To transform the denoised graph spectrum $\mathbf{\hat{S}}_{\mathcal{G}}$ back to the time domain, the inverse operation requires $\boldsymbol{\Psi}_{\mathcal{F}}^{-1}$. Instead of explicit matrix inversion, we propose using 1-D Convolution Network to learn the inversion process $\boldsymbol{\Psi}_{\mathcal{F}}^{-1}$, mitigating numerical instability in the GFT-SVD framework. This trainable module learns the inverse transformation via:
\[
\text{iGFT-SVD-CNet-LA}(\boldsymbol{\Psi}_{\mathcal{F}}) = \text{CNet}_{\theta}(\boldsymbol{\Psi}_{\mathcal{F}})
\]
where $\theta$ denotes network parameters. The inverse graph Fourier transform version of $\hat{\textbf{S}}_G$ using the inversion of GFT-SVD-LA with CNet (iGFY-SVD-CNet-LA) can be formulated as:
\textcolor{black}{
\begin{equation}
\hat{\textbf{s}} = \text{iGFT-SVD-CNet-LA}({{\bf \hat{S}}}_{{\cal{G}}})
=\text{CNet}_{\theta}({{\boldsymbol \Psi}_{}}_{\cal F}){{\bf \hat{S}}_{\cal{G}}}  
\end{equation}}
\textcolor{black}{We integrate the proposed iGFT-SVD-CNet-LA into the decoder module of neural speech enhancement architectures. This replaces conventional Fourier-based processing, improving joint magnitude-phase alignment modeling in monaural speech enhancement.}

\begin{figure}[!t]
\centering
\begin{subfigure}[t]{0.48\columnwidth}
\captionsetup{justification=centering}
\centerline{\includegraphics[width=\columnwidth]{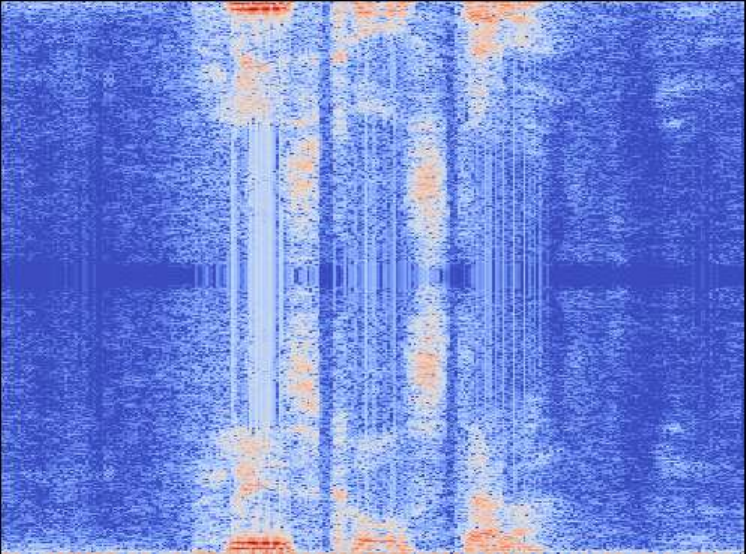}}
\caption{real parts of STFT}
\end{subfigure}
\begin{subfigure}[t]{0.48\columnwidth}
\centerline{\includegraphics[width=\columnwidth]{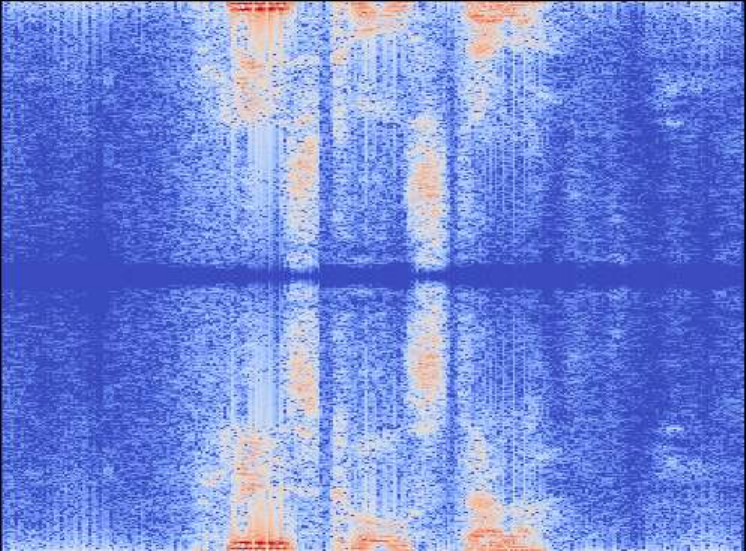}}
\caption{imaginary parts of STFT}
\end{subfigure}
\vskip 0.1in
\begin{subfigure}[t]{0.48\columnwidth}
\centerline{\includegraphics[width=\columnwidth]{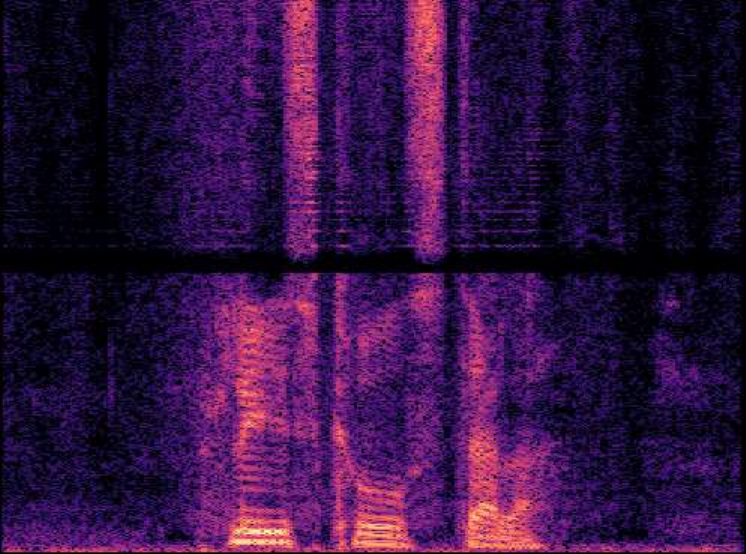}}
\caption{graph spectrum ($p=1\%$)}
\end{subfigure}
\begin{subfigure}[t]{0.48\columnwidth}
\centerline{\includegraphics[width=\columnwidth]{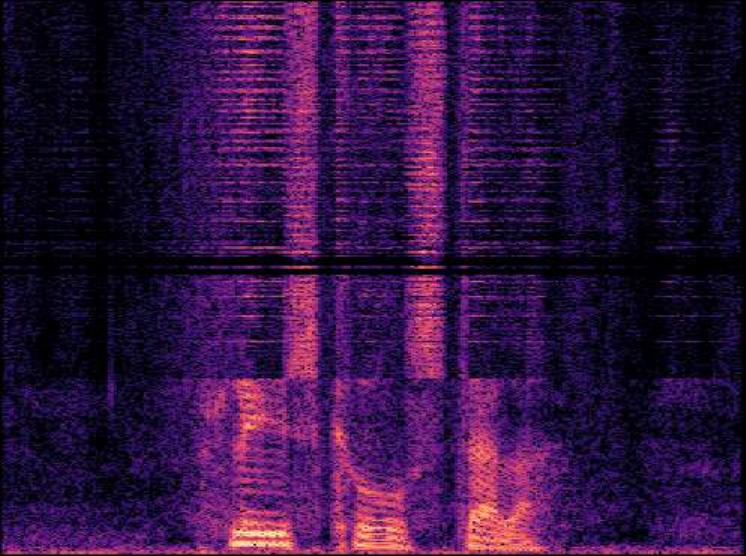}}
\caption{graph spectrum ($p=12\%$)}
\end{subfigure}
\vskip 0.1in
\begin{subfigure}[t]{0.48\columnwidth}
\centerline{\includegraphics[width=\columnwidth]{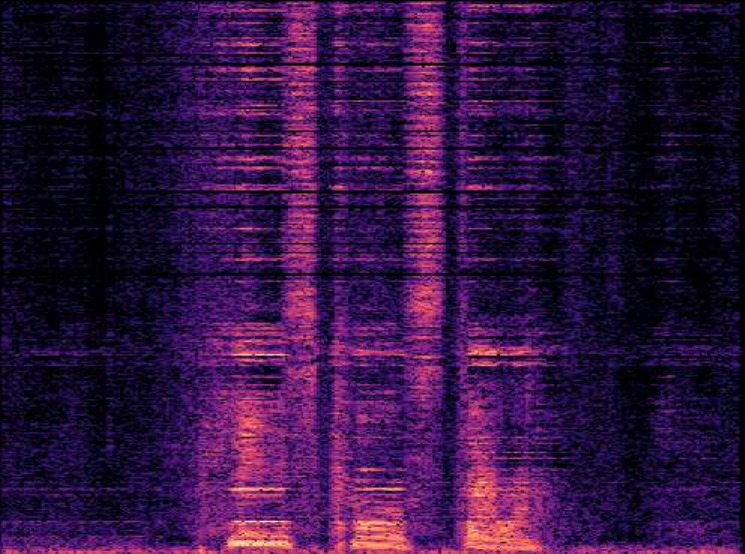}}
\caption{graph spectrum ($p=20\%$)}
\end{subfigure}
\begin{subfigure}[t]{0.48\columnwidth}
\centerline{\includegraphics[width=\columnwidth]{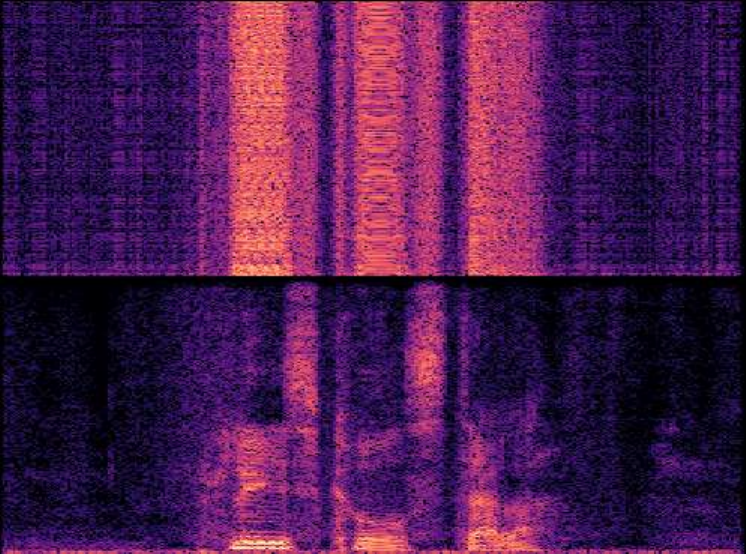}}
\caption{graph spectrum ($p=100\%$)}
\end{subfigure}
\caption{\textcolor{black}{Visualization of a clean speech in the STFT and graph spectrum representations. The (a) real and (b) imaginary parts of the clean STFT spectrum, and the real-valued graph spectrum learned by GFT-SVD-LA with different graph sparsity levels $p$, i.e., (c) $p=1\%$, (d) $p=12\%$, (e) $p=20\%$, and (f) $p=100\%$.}}
\label{fig:spectrm1}
\end{figure}

\textcolor{black}{In Fig.~\ref{fig:spectrm1}, we illustrates the real and imaginary components of an example clean speech signal (from Voice Bank Corpus) obtained using the STFT, and the graph spectra generated by the GFT-SVD-LA under different initial graph sparsity levels $p=1\%, 4\%, 12\%, 20\%, 40\%, 100\%$. It can be observed that the proposed GFT-SVD-LA produces an asymmetric spectrum, thereby avoiding the strict symmetry constraint typically imposed in STFT. This enhances the expressiveness for non-stationary speech signals. Meanwhile, STFT uses a fixed frequency resolution across the entire spectrum, whereas human auditory perception exhibits non-uniform frequency resolution—higher in low-frequency regions and lower in high-frequency ones. In contrast, varying graph sparsity levels reconfigure various graph adjacency matrices, meaning that the time-graph domain can separate the useful speech from the useless noise in different degrees. The proposed GFT-SVD-LA offers the flexibility to define graph topologies that better model perceptual frequency scales. Moreover, a higher $p$ strengthens the capacity of graph topology to capture relationships across noisy speech samples, enriching real-valued spectral details in clean signals. This enhances the accuracy for subsequently enabling mask estimation.}





\section{Experimental Setup} 
\label{sec:setups}
\subsection{Benchmarks and Datasets} 
\textcolor{black}{We perform comprehensive evaluation experiments on two commonly used benchmarks in speech enhancement field, i.e., 2020 Deep Noise Suppression Challenge (DNS-2020)~\cite{DNS2020} and VCTK+DEMAND benchmarks~\cite{voicebank}.} 

\textcolor{black}{\textbf{DNS-2020.} The DNS-2020 benchmark comprises $500$ hours of clean speech data from $2\,150$ speakers and $180$ hours of noise recordings across $150$ noise classes. For training, $72\,000$ 5-second noisy mixtures ($100$ hours in total) are generated by mixing a randomly selected clean utterance from the LibriVox corpus with a randomly selected noise segment at an SNR level randomly sampled from –5 dB to 15 dB in 1 dB increments. Following the setup in~\cite{LvHZX21}, the training and validation sets are split from the generated mixtures using a 4:1 ratio. \textcolor{black}{We use the DNS-2020 blind test set (without reverberation) for evaluation.}}

\textcolor{black}{\textbf{VCTK+DEMAND.} The clean speech data is drawn from the Voice Bank corpus~\cite{VCTK+DEMAND}, which contains $12\,399$ utterances spoken by $30$ speakers. Among them, $11\,572$ and $827$ utterances are used for training and testing, respectively. The noise data includes $8$ real-word noise types from the Diverse Environments Multichannel Acoustic Noise Database (DEMAND)~\cite{DEMAND} and $2$ artificial noise types. The noisy mixtures in training set are generated by mixing the clean speech utterances with the noise clips at SNRs of $0$, $5$, $10$, and $15$ dB. The SNR levels for testing are $2.5$, $7.5$, $12.$5, and $17.5$ dB. Both speakers and noise types for testing are unseen during training.
}

\subsection{\textcolor{black}{Backbone Network}}
\label{subsec:dvc}
\textcolor{black}{To demonstrate the efficacy of the learnable graph topology and GFT-SVD, our proposed GFT-SVD-LA and iGFT-SVD-CNet-LA are systematically evaluated on several well-studied speech enhancement backbone network models. These include \textcolor{black}{both causal and non-causal models}, i.e., Noise Suppression Network (NSNet)~\cite{NSNet}, Dual-Path Convolution Recurrent Network (DPCRN)~\cite{DPCRN}, Dense Convolution Recurrent Network (DCRN)~\cite{DCRN}, Deep Complex Convolution Recurrent Network (DCCRN)~\cite{DCCRN}, Multi-Scale Temporal Frequency Convolution Network with Axial Self-Attention~(MTFAA-Net)~\cite{MTFAA}, UNet with GFT (G-UNet)~\cite{ZhangP22}, Band-Split RNN (BSRNN)~\cite{bsrnn}, \textcolor{black}{Convolutional Time-domain Audio Separation Network (Conv-TasNet)~\cite{LuoM19}, and Dual-Path Transformer Network (DPTNet)~\cite{DPTNet}.}} 

\textcolor{black}{The following provides a brief description of these backbone networks.}
\begin{itemize}
    \item \textcolor{black}{NSNet~\cite{NSNet} is used as the baseline system in Deep Noise Suppression (DNS) Challenge 2020. It is composed of three stacked GRU layers followed by a fully-connected (FC) layer with sigmoid activation to estimate the spectral magnitude gain, where fixed-weighted and SNR-weighted coefficients are introduced in loss function for separate control of speech distortion and noise reduction.}  
    \item \textcolor{black}{DPCRN~\cite{DPCRN} consists of a convolutional encoder, stacked two dual-path RNN (DPRNN) layers, and a decoder composed of transposed convolutional layers. Each DPRNN layer includes an intra-chunk and inter-chunk BLSTM.}
    \item \textcolor{black}{Similar to DPCRN, DCRN~\cite{DCRN} consists of an encoder with dense-connected convolution layers, a decoder, and two stacked BLSTM layers between encoder and decoder for temporal context dependencies modeling.}
    \item \textcolor{black}{DCCRN~\cite{DCCRN} adopts the convolutional encoder-decoder (CED) architecture of DCRN~\cite{DCRN}, but replaces the real-valued CNN and LSTM layers with their complex-valued counterparts to better model the magnitude-phase correlations.}
    \item \textcolor{black}{MTFAA-Net~\cite{MTFAA} is composed of a phase encoder, band merging and band splitting modules, Main-Net module, and mask estimating module. The Main-Net incorporates multi-scale temporal frequency processing and streaming axial self-attention to model the long-range dependencies.}
    \item \textcolor{black}{G-UNet~\cite{ZhangP22} employs an UNet model with an encoder-LSTM-decoder structure to estimate an ideal time-graph ratio mask (IGRM) given the noisy time-graph representations, which are obtained using the Graph Fourier Transform (GFT) based on the graph Laplacian matrix.}   
    \item \textcolor{black}{BSRNN~\cite{bsrnn} involves a band split module, a band and sequence modeling module, and a mask estimation module. It employs bi- and uni-directional band-level modeling for the low- and high-frequency components, respectively, to mitigate the negative impact of unstable high-frequency components.}
    \item \textcolor{black}{Conv-TasNet~\cite{LuoM19} directly learns a latent representation from the noisy waveform using a 1-D convolution (Conv-1D) layer, which is then fed into temporal convolutional network (TCN) for mask estimation. The clean waveform is reconstructed from the masked latent representation using a decoder.}
    \item \textcolor{black}{DPTNet~\cite{DPTNet} shares a CED structure similar to DPCRN and DCCRN, featuring densely-connected convolutional blocks in the encoder and decoder, and stacked dual-path Transformer blocks for temporal dependencies modeling.} 
\end{itemize}

\subsection{Loss Function} 

\textcolor{black}{For a fair and consistent comparison across the backbone models operating in different domains (e.g., STFT and GFT domains), we use same loss function for training. To this end, we adopt the SI-SDR loss~\cite{lossSISDR}, which is computed in the time domain and thus provides a unified optimization objective regardless of the model's feature domain. The SI-SDR loss is defined as follows:
\begin{equation}
    \label{eqa:loss_sisnr}
    \mathcal{L}_{\text{SI-SDR}} = 20 \log_{10} \frac{||\frac{<\hat{\mathbf{s}},{\mathbf{s}}>{\mathbf{s}}}{||{\mathbf{s}}||^2}||}{||\hat{\mathbf{s}} - \frac{<\hat{\mathbf{s}},{\mathbf{s}}>{\mathbf{s}}}{||{\mathbf{s}}||^2}||}
\end{equation}
where $\hat{\mathbf{s}}$ and $\mathbf{s}$ denote the estimated and clean speech waveform, respectively. The notation ${<,>}$ denotes the dot product operation between two vectors, and $||.||$ denotes the Euclidean (L2) norm. Note that $||\hat{\mathbf{s}} - \frac{\langle \hat{\mathbf{s}}, \mathbf{s} \rangle \mathbf{s}}{||\mathbf{s}||^2}||$ inherently captures phase differences through the geometry of projection. This enables SI-SDR to function as an implicit phase-aware loss, penalizing phase misalignment through waveform orthogonality, without relying on explicit short-time spectral phase~\cite{FanSisnr,LuoM19}.}
\subsection{\textcolor{black}{Implementation Details}}
\textcolor{black}{All audio recordings are sampled at a frequency of 16 kHz. The window of length $25$ ms with a hop length of $6.25$ ms is used for spectral analysis. A 512-point GFT is applied for each time frame, leading to a 512-point time-graph spectrum as the input to models. For a fair comparison, 512-point STFT spectrum is calculated in a same way as the input to models. We keep the default parameter configurations across the baseline backbone models.}

\textcolor{black}{All the models are trained using PyTorch on an NVIDIA GeForce RTX 3090 Ti GPU. A mini-batch size of $4$ speech utterances is used for each training iteration. All models are trained using the Adam optimizer with an initial learning rate of $1 \times 10^{-3}$. Early stopping is employed, with training halted if the validation loss does not improve for 20 consecutive epochs.} \textcolor{black}{The decay factor $\gamma=0.75$ is used to scale the learning rate if the validation metric does not improve for 5 consecutive epochs.} 
\subsection{Evaluation Metrics}
\textcolor{black}{In our experiments, three objective metrics are adopted for performance evaluation, including the perceptual evaluation of speech quality (PESQ)~\cite{PESQ} for speech quality, short-time objective intelligibility (STOI)~\cite{STOI} for speech intelligibility, and scale-invariant signal-to-distortion ratio (SI-SDR)~\cite{lossSISDR} for speech distortion. Following prior studies~\cite{bsrnn, LiYYZL22}, we report both narrow-band (N-PESQ) and wide-band PESQ (W-PESQ) results. For all the metrics, higher scores indicate better performance.}

\begin{table}[h]
    \centering
    \def\arraystretch{1.2}
    \setlength{\tabcolsep}{2pt}
    \setlength{\abovetopsep}{0pt}
    \setlength\belowbottomsep{0pt} 
    \setlength\aboverulesep{0pt} 
    \setlength\belowrulesep{0pt}
  \caption{\textcolor{black}{The \textcolor{black}{W-PESQ evaluation} results for different graph topologies across various graph sparsity levels ($p$), on the VCTK+DEMAND benchmark.}}
  \begin{tabular}{l|cccccc}
  \toprule[1.1pt] 
  \multirow{2}{*}{\bf Graph Topology Type} & \multicolumn{6}{c}{\bf Graph Sparsity ($p$)} \\
  & \bf 1\% & \bf 4\% &  {\bf 12\%} & \bf 20\% & \bf 40\%  &  {\bf 100\%} \\
  \midrule
  \textcolor{black}{Fixed Graph Topology~\cite{timegraph} }& \textcolor{black}{2.462} & \textcolor{black}{2.447}  & \textcolor{black}{2.279}  & \textcolor{black}{2.415} &\textcolor{black}{2.446} & \textcolor{black}{ 2.167}\\
     \textcolor{black}{Neural Graph Topology~\cite{ZhangP22} }& \textcolor{black}{ 2.525} & \textcolor{black}{2.480} &   \textcolor{black}{2.362} &  \textcolor{black}{2.474} & \textcolor{black}{2.511} & \textcolor{black}{2.213} \\
    \textcolor{black}{Learnable Graph Topology} & \textcolor{black}{2.607} & \textcolor{black}{2.551} & \textcolor{black}{2.487} & \textcolor{black}{2.548}  & \textcolor{black}{2.533} & \textcolor{black}{2.512} \\
    \toprule[1.1pt] 
  \end{tabular}
  \label{table:k_params}
\end{table}

\section{Results and discussions }\label{results}
\subsection{Graph Sparsity}
\textcolor{black}{In this section, we first evaluate the efficacy of the learnable graph topology across various graph sparsity levels $p$, benchmarking against fixed~\cite{timegraph} and neural graph topology~\cite{ZhangP22}.} From Table~\ref{table:k_params}, we can observe that higher $p$ values increase computational complexity without commensurate performance gains, \textcolor{black}{while insufficient sparsity fails to capture the intrinsic structure of speech samples. The choice of $p=1\%$ yields optimal results in both topological capacity and computational efficiency, aligning with the findings in previous studies~\cite{YanYWG20, ref9}.} The proposed learnable graph topology achieves superior quality scores across all sparsity levels, demonstrating the efficacy of graph shift operator initialization for graph adjacency matrices.
\begin{table}[!t]
	\caption{\textcolor{black}{Performance comparison of the backbone models using GFT-SVD-LA, GFT-SVD-CNet, GFT-SVD-CNet-LA on the VCTK+DEMAND benchmark, evaluated in terms of N-PESQ, W-PESQ, SI-SDR (dB), and STOI (in \%).}}
    \def\arraystretch{1.1}
    \setlength{\tabcolsep}{2.0pt}
  \begin{tabular}{lcccccccc}
    \toprule[1.0pt] 
   \bf{Network}&  \bf GFT &  \multirow{1}*{\bf{W-PESQ}}  & \multirow{1}*{\bf{N-PESQ}}  & \bf{SI-SDR}  & \multirow{1}*{\bf{STOI}}  \\
   \midrule \midrule 
    {NSNet-LA$_{\cal {G}}$} & SVD-LA &{2.525}&{3.275}& {17.768}& {93.2}\\ 
    {NSNet-CNet$_{\cal {G}}$} & SVD-CNet &{2.556}&{3.289}&{17.802}&{93.3}\\ 
    {NSNet-CL$_{\cal {G}}$} & SVD-CNet-LA &{2.607}&{3.298}& {17.471}& {93.3}\\ 
   \midrule
    {DPCRN-LA$_{\cal {G}}$} & SVD-LA&{2.807}&{3.567}& {17.768}& {93.2}\\ 
    {DPCRN-CNet$_{\cal {G}}$} & SVD-CNet&{2.810}&{3.596}& {18.874}& {94.3}\\ 
    {DPCRN-CL$_{\cal {G}}$} & SVD-CNet-LA&{2.820}&{3.590}& {19.077}& {94.3}\\ 
   \midrule
    {DCRN-LA$_{\cal {G}}$} &SVD-LA&{2.674} &{3.448}& {18.618 }& {93.7} \\ 
    {DCRN-CNet$_{\cal {G}}$} &SVD-CNet&{2.705} &{3.478}& {18.479}& {93.7} \\ 
    {DCRN-CL$_{\cal {G}}$} &SVD-CNet-LA&{2.723} &{3.495}& {18.372}& {93.7} \\ 
  \midrule
    {DCCRN-LA$_{\cal {G}}$} &SVD-LA & {2.713}& {3.490}& {18.526}& {93.6}\\
    {DCCRN-CNet$_{\cal {G}}$} &SVD-CNet & {2.723}& {3.508}& {18.181}& {93.8}\\
    {DCCRN-CL$_{\cal {G}}$} &SVD-CNet-LA & {2.720}& {3.516}& {18.349}& {93.7}\\
  \midrule
	{MTFAA-LA$_{\cal {G}}$} & SVD-LA& 2.750&3.512& 18.807& 93.8\\
    {MTFAA-CNet$_{\cal {G}}$} & SVD-CNet&2.766 &3.526 &18.802 & 93.8\\
    {MTFAA-CL$_{\cal {G}}$} & SVD-CNet-LA&2.763 &3.538&18.802 & 93.5\\
  \midrule
   {BSRNN-LA$_{\cal {G}}$} &SVD-LA & 2.460 & 3.283 & 18.207 & 92.3\\ 
   {BSRNN-CNet$_{\cal {G}}$} &SVD-CNet & 2.523 & 3.299 & 17.832 & 92.8\\ 
   {BSRNN-CL$_{\cal {G}}$} &SVD-CNet-LA & {2.533}&{3.313}& {18.524}& 93.0 \\ 
   \midrule
   {UNet-LA}$_{\cal {G}}$ &SVD-LA &{2.731} &{3.526}&{18.092} &{94.0} \\ 
   {{UNet-CNet}$_{\cal {G}}$} &SVD-CNet &{2.752} &{3.530}&{18.164} &{93.8} \\ 
   {{UNet-CL}$_{\cal {G}}$} &SVD-CNet-LA &{2.734} &{3.533}&{18.278} &{93.6} \\ 
   \midrule 
   {\textcolor{black}{ Conv-TasNet-LA}$_{\cal {G}}$} &\textcolor{black}{SVD-LA} &{\textcolor{black}{2.541}} &{\textcolor{black}{3.346}}&{\textcolor{black}{16.995}} &{\textcolor{black}{93.5}} \\ 
   {\textcolor{black}{ Conv-TasNet-CNet}$_{\cal {G}}$} &\textcolor{black}{SVD-CNet} &{\textcolor{black}{2.608}} &{\textcolor{black}{3.351}}&{\textcolor{black}{17.603}} &{\textcolor{black}{93.6}} \\ 
   {\textcolor{black}{ Conv-TasNet-CL}$_{\cal {G}}$} &\textcolor{black}{SVD-CNet-LA} &{\textcolor{black}{2.617}} &{\textcolor{black}{3.370}}&{\textcolor{black}{17.269}} &{\textcolor{black}{93.4}}  \\   
   \midrule 
   {\textcolor{black}{DPTNet-LA}$_{\cal {G}}$} &\textcolor{black}{SVD-LA} &{\textcolor{black}{2.788}} &{\textcolor{black}{3.639}}&{\textcolor{black}{19.860}} &{\textcolor{black}{94.7}} \\ 
   {\textcolor{black}{DPTNet-CNet}$_{\cal {G}}$} &\textcolor{black}{SVD-CNet} &{\textcolor{black}{2.842}} &{\textcolor{black}{3.664}}&{\textcolor{black}{19.458}} &{\textcolor{black}{94.6}} \\ 
   {\textcolor{black}{DPTNet-CL}$_{\cal {G}}$} &\textcolor{black}{SVD-CNet-LA} &{\textcolor{black}{2.891}} &{\textcolor{black}{3.680}}&{\textcolor{black}{19.797}} &{\textcolor{black}{94.7}}  \\ 
   \toprule[1.0pt]
  \end{tabular}
  \label{table_VCTK+DEMAND}
\end{table}

\textcolor{black}{\subsection{Ablation Study}
 In this section, we evaluate the our learnable GFT-SVD against GFT-SVD-LA, GFT-SVD-CNet. Table~\ref{table_VCTK+DEMAND} compares the performance of nine backbone networks with each graph Fourier basis variant on the VCTK+DEMAND benchmark, where 
LA-variants represent the combination of backbone networks and GFT-SVD-LA, CNet-variants represent the combination of backbone networks and GFT-SVD-CNet, and CL-variants represent backbone networks with GFT-SVD-CNet-LA. It can be seen that the proposed GFT-SVD-CNet-LA achieves state-of-the-art performance across all backbones. LA-variants demonstrate that graph shift operator initialization ($\mathbf{W}$) provides a theoretically grounded topology. CNet-variants confirm matrix inversion error elimination via Conv-1D 
approximation. Both mechanisms enhance the alignment magnitude and phase modeling in DNNs-based SE methods.}

Table~\ref{DNS 2020 no-reverb DATASET} shows the comparison results of the combination of backbone networks and GFT-SVD-LA, GFT-SVD-CNet, GFT-SVD-CNet-CL on the DNS-2020 no-reverb benchmark. It is observed that the GFT-SVD-CNet-LA configuration consistently outperforms baselines with GFT-SVD-LA, GFT-SVD-CNet. These cross-dataset results 
confirm that the simultaneous optimization of learnable graph topology and learnable graph Fourier basis is critical for high-fidelity speech enhancement.
\begin{table}[!t]
	\caption{\textcolor{black}{Performance comparison of the backbone models using GFT-SVD-LA, GFT-SVD-CNet, GFT-SVD-CNet-LA on the DNS 2020 no-reverb benchmark, evaluated in terms of N-PESQ, W-PESQ, SI-SDR (dB), and STOI (in \%).}}
    \def\arraystretch{1.1}
    \setlength{\tabcolsep}{2.0pt}
  \begin{tabular}{lcccccccc}
    \toprule[1.3pt] 
   \bf{Network}&  \bf GFT &  \multirow{1}*{\bf{W-PESQ}}  & \multirow{1}*{\bf{N-PESQ}}  & \bf{SI-SDR}  & \multirow{1}*{\bf{STOI}}
   \\\midrule \midrule 
   {NSNet-LA$_{\cal {G}}$} & SVD-LA &{2.475}&{3.023}& {16.573}& {95.2}\\
    {NSNet-CNet$_{\cal {G}}$} & SVD-CNet &{2.479}&{3.033}& {16.590}& {95.2}\\ 
      {NSNet-CL$_{\cal {G}}$} & SVD-CNet-LA &{2.588}&{3.135}& {17.317}& {95.8}\\ 
     \midrule
{DPCRN-LA$_{\cal {G}}$} & SVD-LA&{2.938}&{3.434}& {18.959}& {97.0}\\ 
{DPCRN-CNet$_{\cal {G}}$} & SVD-CNet&{2.954}&{3.461}& {19.172}& {97.1}\\
{DPCRN-CL$_{\cal {G}}$} & SVD-CNet-LA&{3.000}&{3.489}& {19.363}& {97.1}\\ 
\midrule
 {DCRN-LA$_{\cal {G}}$} &SVD-LA & {2.754}& {3.272}& {18.268}& {96.3}\\
 {DCRN-CNet$_{\cal {G}}$} &SVD-CNet&{2.773} &{3.282}& {18.294 }& {96.3} \\ 
  {DCRN-CL$_{\cal {G}}$} &SVD-CNet-LA&{2.792} &{3.295}& {18.456}& {96.4} \\ 
  \midrule
{DCCRN-LA$_{\cal {G}}$} &SVD-LA & {2.836}& {3.369}& {18.722}& {96.7}\\
 {DCCRN-CNet$_{\cal {G}}$} &SVD-CNet & {2.861}& {3.389}& {18.933}& {96.7}\\
  {DCCRN-CL$_{\cal {G}}$} &SVD-CNet-LA & {2.875}& {3.414}& {19.004}& {96.8}\\
  \midrule
	{MTFAA-LA$_{\cal {G}}$} &SVD-LA& 2.722 & 3.260& 18.266 & 96.3 \\
    {MTFAA-CNet$_{\cal {G}}$} & SVD-CNet& 2.795 & 3.311& 18.512 & 96.5 \\
   {MTFAA-CL$_{\cal {G}}$} & SVD-CNet-LA& 2.865 & 3.390&18.868 &96.8 \\
  \midrule
    {BSRNN-LA$_{\cal {G}}$} &SVD-LA & {2.197}&{2.739}& {14.844}& 93.2 \\ 
      {BSRNN-CNet$_{\cal {G}}$} &SVD-CNet &2.211 &{2.726}& {14.773}& 93.3  \\ 
         {BSRNN-CL$_{\cal {G}}$} &SVD-CNet-LA & {2.229}&{2.774}& {15.233}& 93.6 \\ 
        \midrule
        {{UNet-LA}$_{\cal {G}}$} &SVD-LA &{2.770} &{3.331}&{18.682} &{96.5} \\ 
         {{UNet-CNet}$_{\cal {G}}$} &SVD-CNet &{2.794} &{3.344}&{18.554} &{96.6} \\ 
        {{UNet-CL}$_{\cal {G}}$} &SVD-CNet-LA &{2.812} &{3.352}&{18.554} &{96.7} \\ 
   \midrule
     {\textcolor{black}{Conv-TasNet-LA}$_{\cal {G}}$} &\textcolor{black}{SVD-LA} &{\textcolor{black}{2.628}} &{\textcolor{black}{3.206}}&{\textcolor{black}{15.596}} &{\textcolor{black}{95.8}} \\ 
         {\textcolor{black}{ Conv-TasNet-CNet}$_{\cal {G}}$} &\textcolor{black}{SVD-CNet} &{\textcolor{black}{2.687}} &{\textcolor{black}{3.249}}&{\textcolor{black}{16.590}} &{\textcolor{black}{96.0}} \\ 
        {\textcolor{black}{ Conv-TasNet-CL}$_{\cal {G}}$} &\textcolor{black}{SVD-CNet-LA}  &{\textcolor{black}{2.707}} &{\textcolor{black}{3.260}}&{\textcolor{black}{16.472}} &{\textcolor{black}{96.1}} \\ 
      \toprule[1.3pt]
  \end{tabular}
  \label{DNS 2020 no-reverb DATASET}
\end{table}

\textcolor{black}{\subsection{Comparison Study}
\textcolor{black}{In Table~\ref{table_Voicebank+Demand}, we compare the proposed GFT-SVD-CNet-LA with STFT, GFT-EVD, GFT-SVD, and Conv-1D on the VCTK+DEMAND benchmark, across different backbones in terms of PESQ, SI-SDR, and STOI. For the implementations of the backbones, we adopt the the authors' source code with default model configurations. The comparison results demonstrates that the GFT-SVD-CNet-LA consistently achieves significant improvements over Conv-1D, STFT, GFT-EVD, and GFT-SVD in all the three metrics. For instance, DPCRN-CL$_G$ improves the STFT-based counterpart DPCRN-STFT by 0.269 in W-PESQ, by 0.533 in SI-SDR and 0.3\% in STOI. Among all the models, overall, the performance ranking is GFT-SVD-CNet-LA $>$ GFT-SVD $>$ STFT.}}

\textcolor{black}{\textcolor{black}{Table~\ref{table_DNSchallengs} reports the comparison results of GFT-SVD-CNet-LA, STFT, GFT-EVD, GFT-SVD, and Conv-1D on the DNS-2020 benchmark (No-reverb). Similar performance trends are observed to those in Table~\ref{table_Voicebank+Demand}. Again, our proposed learnable GFT-SVD-CNet-LA achieve the best results across different backbones, significantly outperforming GFT-EVD and GFT-SVD. These comparisons indicate that learnable GFT-SVD strengthens the GFT’s capacity for amplitude–phase alignment modeling.}}

\begin{table}[!t]
	\caption{\textcolor{black}{Performance comparison of the backbone models using STFT, GFT-SVD, and learnable GFT-SVD on the VCTK+DEMAND benchmark, evaluated in terms of N-PESQ, W-PESQ, SI-SDR (dB), and STOI (in \%).}}
    \def\arraystretch{1.1}
    \setlength{\tabcolsep}{1.0pt}
  \begin{tabular}{lcccccccc}
    \toprule[1.0pt] 
  \bf{ Network }&  \bf Transform& \multirow{1}*{\bf{W-PESQ}}  & \multirow{1}*{\bf{N-PESQ}}  & \bf{SI-SDR}  & \multirow{1}*{\bf{STOI}}
   \\\midrule \midrule 
   	NSNet~\cite{NSNet}& \multirow{1}*{STFT} & 2.343&3.109& {17.793}&{92.8}\\
    	{NSNet$_{\cal {G}}$}~\cite{timegraph} &GFT-SVD&{2.462} &{3.215}&{18.035} &{93.2}\\ 
       {NSNet-CL$_{\cal {G}}$} &GFT-SVD-CNet-LA&{2.607} &{3.298}&{17.471} &{93.3}\\
        \midrule 
		DPCRN~\cite{DPCRN} &STFT & {2.551}&{3.333}& {18.544}& {94.0}\\ 
  {DPCRN$_{\cal {G}}$}~\cite{timegraph} &GFT-SVD &{2.741}&{3.556}& {18.826}&{94.4}\\
  {DPCRN-CL$_{\cal {G}}$} &GFT-SVD-CNet-LA &{2.820}&{3.590}& {19.077}&{94.3}\\ 
  \midrule 
		DCRN~\cite{DCRN} &STFT & 2.436&{3.211}& {17.877}& {93.3} \\
  {DCRN$_{\cal {G}}$}~\cite{timegraph}&GFT-SVD &{2.624}&{3.426}& {18.242}&{93.7}\\
 {DCRN-CL$_{\cal {G}}$}&GFT-SVD-CNet-LA &{2.723}&{3.495}& {18.372}&{93.7}\\
  \midrule 
		DCCRN~\cite{DCCRN} &  STFT& {2.510}&{3.255}& {18.242}&{93.3}\\
 {DCCRN$_{\cal {G}}$}~\cite{timegraph}& GFT-SVD & {2.634}& {3.468}& {18.125}&{94.0} \\
 {DCCRN-CL$_{\cal {G}}$}&GFT-SVD-CNet-LA &{2.720}&{3.516}& {18.349}&{93.7}\\
  \midrule 
		MTFAA~\cite{MTFAA} &STFT& {2.657}&{3.478}& {18.837}&{93.8}\\ 
  {MTFAA$_{\cal {G}}$}~\cite{timegraph}& GFT-SVD&2.748 &3.553&18.670 & 94.2\\
   {MTFAA-CL$_{\cal {G}}$}& GFT-SVD-CNet-LA&2.763 &3.538&18.802 & 93.5\\
  \midrule 
        {BSRNN}~\cite{bsrnn}&STFT& {2.444}&{3.190}& {18.831}&92.8  \\
        {BSRNN$_{\cal {G}}$}~\cite{timegraph} & GFT-SVD&{2.567}&{3.358}& {19.219}& {93.2}\\ 
        {BSRNN-CL$_{\cal {G}}$} & GFT-SVD-CNet-LA&{2.533}&{3.313}&{18.524}&{93.0}\\ 
        \midrule 
        UNet~\cite{Unet} & \multirow{1}*{STFT} & {2.160}&{3.000}& {16.484}&92.6   \\
         G-UNet~\cite{ZhangP22}&GFT-EVD &2.455 & 3.274&18.009 &93.2 \\ 
       {UNet$_{\cal {G}}$}~\cite{timegraph}& GFT-SVD &{2.644} &{3.490}&{18.611} &{93.7} \\ 
        {UNet-CL$_{\cal {G}}$}& GFT-SVD-CNet-LA &{2.734} &{3.533}&{18.278} &{93.6} \\ 
        \midrule 
        \textcolor{black}{{{Conv-TasNet}}~\cite{LuoM19}} & \textcolor{black}{Conv-1D} &{\textcolor{black}{2.537}} &{\textcolor{black}{3.318}}&{\textcolor{black}{19.178}} &{\textcolor{black}{93.6}} \\
       {\textcolor{black}{{Conv-TasNet}}} & \textcolor{black}{STFT} &{\textcolor{black}{2.328}} &{\textcolor{black}{3.067}}&{\textcolor{black}{16.825}} &{\textcolor{black}{92.5}} \\
         {\textcolor{black}{{Conv-TasNet}$_{\cal {G}}$}} & \textcolor{black}{GFT-SVD} &{\textcolor{black}{2.532}} &{\textcolor{black}{3.348}}&{\textcolor{black}{16.835}} &{\textcolor{black}{93.5}} \\
        \textcolor{black}{{{Conv-TasNet-CL}$_{\cal {G}}$}} & \textcolor{black}{GFT-SVD-CNet-LA} &{\textcolor{black}{2.617}} &{\textcolor{black}{3.370}}&{\textcolor{black}{17.269}} &{\textcolor{black}{93.4}} \\ 
         \midrule
         {\textcolor{black}{{DPTNet}~\cite{DPTNet}}} & \textcolor{black}{STFT} &{\textcolor{black}{2.538}} &{\textcolor{black}{3.324}}&{\textcolor{black}{18.970}} &{\textcolor{black}{94.0}} \\
         {\textcolor{black}{{DPTNet}$_{\cal {G}}$}} & \textcolor{black}{GFT-SVD}  &{\textcolor{black}{2.875}} &{\textcolor{black}{3.658}}&{\textcolor{black}{19.979}} &{\textcolor{black}{94.8}} \\
        {\textcolor{black}{{DPTNet-CL}$_{\cal {G}}$}} & \textcolor{black}{GFT-SVD-CNet-LA} &{\textcolor{black}{2.891}} &{\textcolor{black}{3.680}}&{\textcolor{black}{19.797}} &{\textcolor{black}{94.7}}  \\ 
      \toprule[1.0pt]
  \end{tabular}
  \label{table_Voicebank+Demand}
\end{table}

\begin{table}
	\caption{\textcolor{black}{Performance comparison of the backbone models using STFT, GFT-SVD, and learnable GFT-SVD on the DNS-2020 benchmark, evaluated in terms of N-PESQ, W-PESQ, SI-SDR (dB), and STOI (in \%).}}
    \def\arraystretch{1.1}
    \setlength{\tabcolsep}{0.8pt}
  \begin{tabular}{lcccccccc}
    \toprule[1.0pt] 
   \bf{Network}&  \bf Transform&  \multirow{1}*{\bf{W-PESQ}}  & \multirow{1}*{\bf{N-PESQ}}  & \bf{SI-SDR}  & \multirow{1}*{\bf{STOI}}  
   \\\midrule \midrule 
   	NSNet~\cite{NSNet}& \multirow{1}*{STFT}  & 2.331&2.865& {16.262}&{94.6}\\
     {NSNet$_{\cal {G}}$}~\cite{timegraph} & GFT-SVD &{2.439}&{3.008}& {16.352}& {95.0}\\ 
      {NSNet-CL$_{\cal {G}}$} & GFT-SVD-CNet-LA &{2.588}&{3.135}& {16.700}& {95.8}\\ 
     \midrule
		DPCRN~\cite{DPCRN}& STFT & {2.797}&{3.260}& {18.570}& {96.7}\\ 
{DPCRN$_{\cal {G}}$}~\cite{timegraph} & GFT-SVD&{2.885}&{3.392}& {18.964}& {96.8}\\ 
{DPCRN-CL$_{\cal {G}}$} & GFT-SVD-CNet-L&{3.000}&{3.489}& {19.363}& {97.1}\\ 
\midrule
		DCRN~\cite{DCRN}& STFT & 2.566&{3.074}& {17.326}& {95.8} \\ 
  {DCRN$_{\cal {G}}$}~\cite{timegraph} &GFT-SVD&{2.739} &{3.272}& {18.143 }& {96.2} \\
 {DCRN-CL$_{\cal {G}}$} &GFT-SVD-CNet-LAA&{2.792} &{3.295}& {18.456 }& {96.4} \\ 
  \midrule
		DCCRN~\cite{DCCRN}& STFT & {2.644}&{ 3.148}& {17.950}&{96.3}\\
 {DCCRN$_{\cal {G}}$}~\cite{timegraph} &GFT-SVD & {2.841}& {3.376}& {18.587}& {96.7}\\
 {DCCRN-CL$_{\cal {G}}$} &GFT-SVD-CNet-LA& {2.875}& {3.414}& {19.004}& {96.8}\\
  \midrule
		MTFAA~\cite{MTFAA}& STFT & {2.696}&{3.243}& {18.294}&{96.4}\\ 
 {MTFAA$_{\cal {G}}$}~\cite{timegraph} & GFT-SVD& 2.707&3.261& 18.454& 96.3\\
   {MTFAA-CL$_{\cal {G}}$} & GFT-SVD-CNet-LA&2.865 &3.390 &18.868 &96.8\\
  \midrule
        BSRNN~\cite{bsrnn}& STFT & {2.203}&{2.734}& {16.070}&94.2  \\
       {BSRNN$_{\cal {G}}$}~\cite{timegraph} &GFT-SVD & {2.321}&{2.835}& {16.300}&94.3  \\ 
        {BSRNN-CL$_{\cal {G}}$} &GFT-SVD-CNet-LA & {2.229}&{2.774}& {15.233}&93.6  \\ 
        \midrule
        UNet~\cite{Unet}  &\multirow{1}*{STFT}& {2.017}&{2.597}& {14.585}&94.0   \\
        G-UNet~\cite{ZhangP22}  &GFT-EVD &2.580 & 3.129&17.583 &95.9 \\       
        {{UNet}$_{\cal {G}}$}~\cite{timegraph} &GFT-SVD &{2.785} &{3.346}&{18.466} &{96.6} \\ 
        {{UNet-CL}$_{\cal {G}}$} &GFT-SVD-CNet-LA &{2.812} &{3.352}&{18.554} &{96.7} \\
          \midrule 
           \textcolor{black}{{{Conv-TasNet}}~\cite{LuoM19}} & \textcolor{black}{Conv-1D} &{\textcolor{black}{2.328}} &{\textcolor{black}{2.866}}&{\textcolor{black}{17.045}} &{\textcolor{black}{94.5}} \\
    {\textcolor{black}{{Conv-TasNet}}} &\textcolor{black}{STFT} &{\textcolor{black}{{2.423}}} &{\textcolor{black}{{2.982}}}&{\textcolor{black}{{15.930}}} &{\textcolor{black}{{95.2}}} \\
        \textcolor{black}{ {{Conv-TasNet}$_{\cal {G}}$}} & \textcolor{black}{GFT-SVD}&{\textcolor{black}{{2.637}}} &{\textcolor{black}{{3.222}}}&{\textcolor{black}{{15.351}}} &{\textcolor{black}{{95.7}}} \\
        \textcolor{black}{{{Conv-TasNet-CL}$_{\cal {G}}$}} &\textcolor{black}{GFT-SVD-CNet-LA}  &{\textcolor{black}{{2.719}}} &{\textcolor{black}{{3.267}}}&{\textcolor{black}{{16.367}}} &{\textcolor{black}{{96.0}}} \\ 
          \midrule
         {\textcolor{black}{{DPTNet}~\cite{DPTNet}}} &\textcolor{black}{STFT} &{\textcolor{black}{2.577}} &{\textcolor{black}{3.095}}&{\textcolor{black}{18.408}} &{\textcolor{black}{95.7}} \\
         {\textcolor{black}{{DPTNet}$_{\cal {G}}$}} &\textcolor{black}{GFT-SVD} &{\textcolor{black}{2.741}} &{\textcolor{black}{3.284}}&{\textcolor{black}{19.076}} &{\textcolor{black}{96.5}} \\
        {\textcolor{black}{{DPTNet-CL}$_{\cal {G}}$}} &\textcolor{black}{GFT-SVD-CNet-LA} &{\textcolor{black}{2.789}} &{\textcolor{black}{3.347}}&{\textcolor{black}{18.875}} &{\textcolor{black}{96.3}}  \\ 
      \toprule[1.0pt]
  \end{tabular}
  \label{table_DNSchallengs}
\end{table}

\begin{figure}[t]
\centering
\begin{subfigure}[t]{0.48\columnwidth}
\captionsetup{justification=centering}
\centerline{\includegraphics[width=\columnwidth]{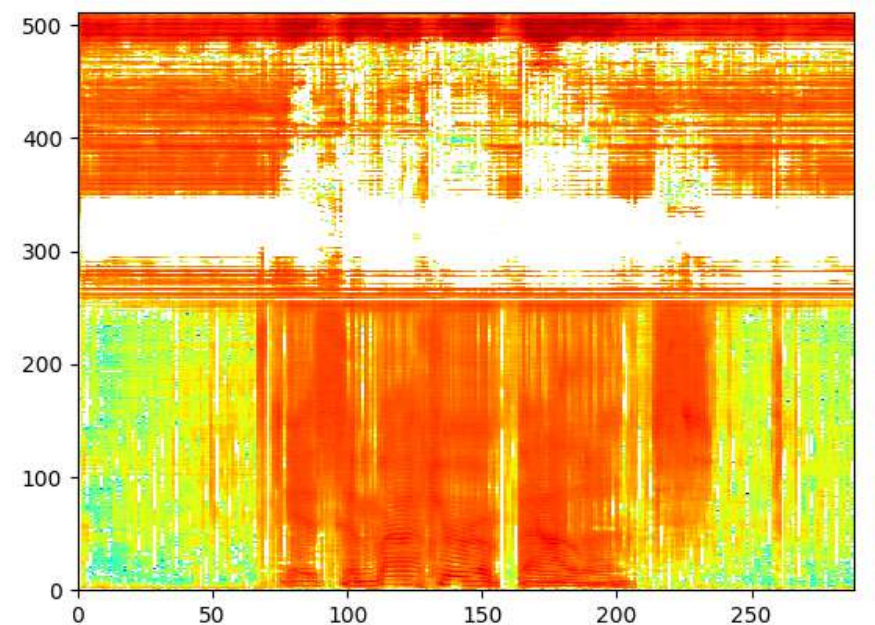}}
\caption{Mask.re (STFT)}
\end{subfigure}
\begin{subfigure}[t]{0.48\columnwidth}
\centerline{\includegraphics[width=\columnwidth]{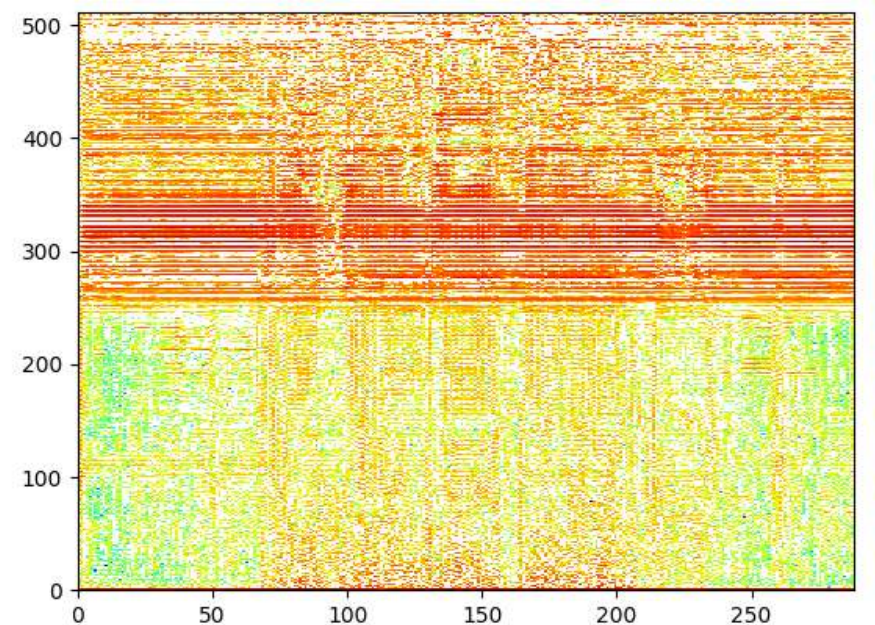}}
\caption{Mask.im (STFT)}
\end{subfigure}
\vskip 0.1in
\begin{subfigure}[t]{0.48\columnwidth}
\centerline{\includegraphics[width=\columnwidth]{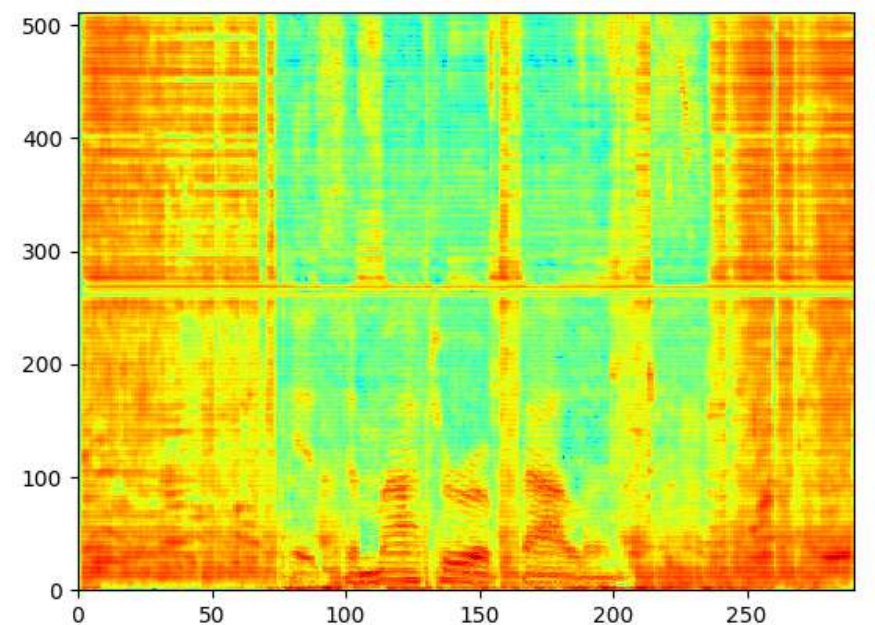}}
\caption{Mask.$1_{th}$ (GFT-SVD-LA)}
\end{subfigure}
\begin{subfigure}[t]{0.48\columnwidth}
\centerline{\includegraphics[width=\columnwidth]{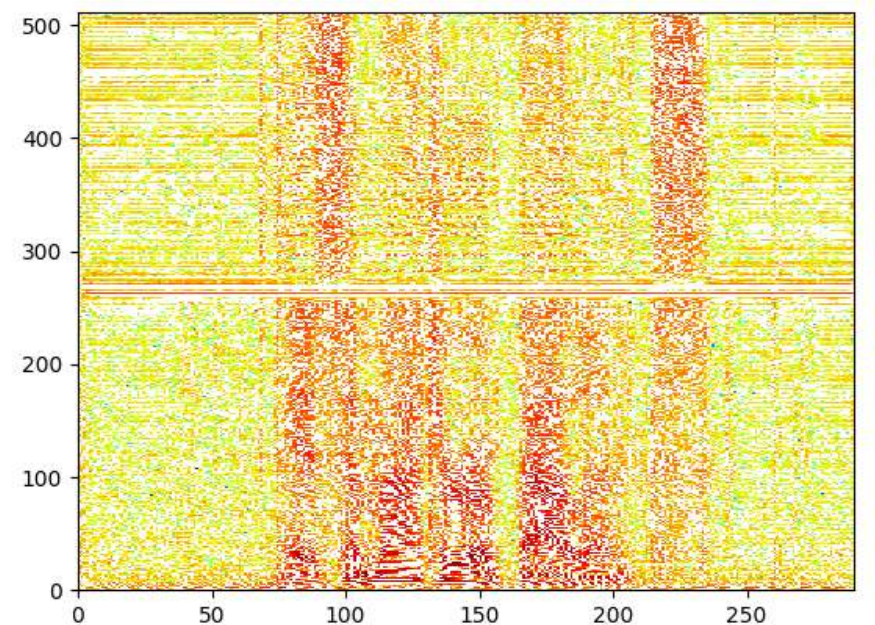}}
\caption{Mask.$2_{th}$ (GFT-SVD-LA)}
\end{subfigure}
\vskip 0.1in
\begin{subfigure}[t]{0.48\columnwidth}
\centerline{\includegraphics[width=\columnwidth]{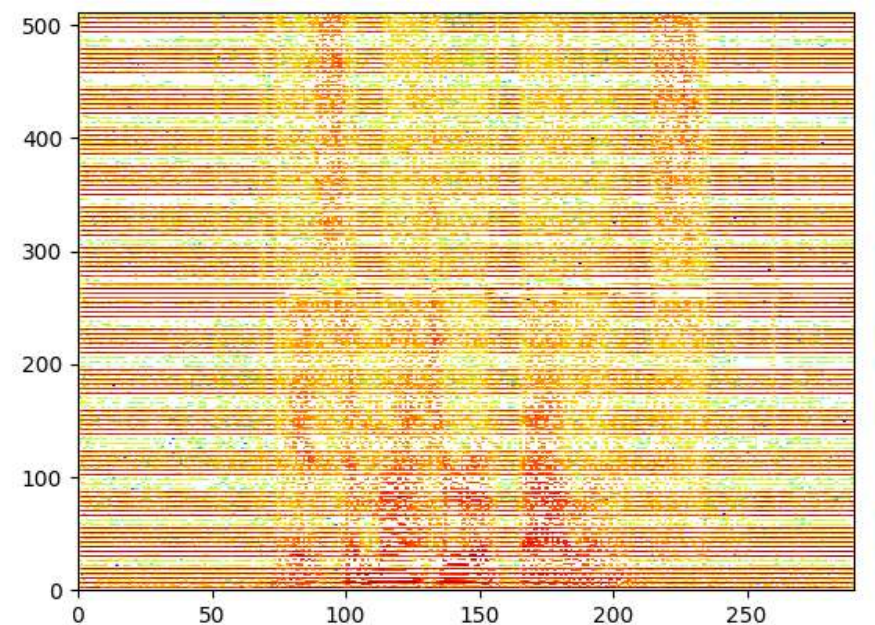}}
\caption{Mask.$1_{th}$ (GFT-SVD-CNet)}
\end{subfigure}
\begin{subfigure}[t]{0.48\columnwidth}
\centerline{\includegraphics[width=\columnwidth]{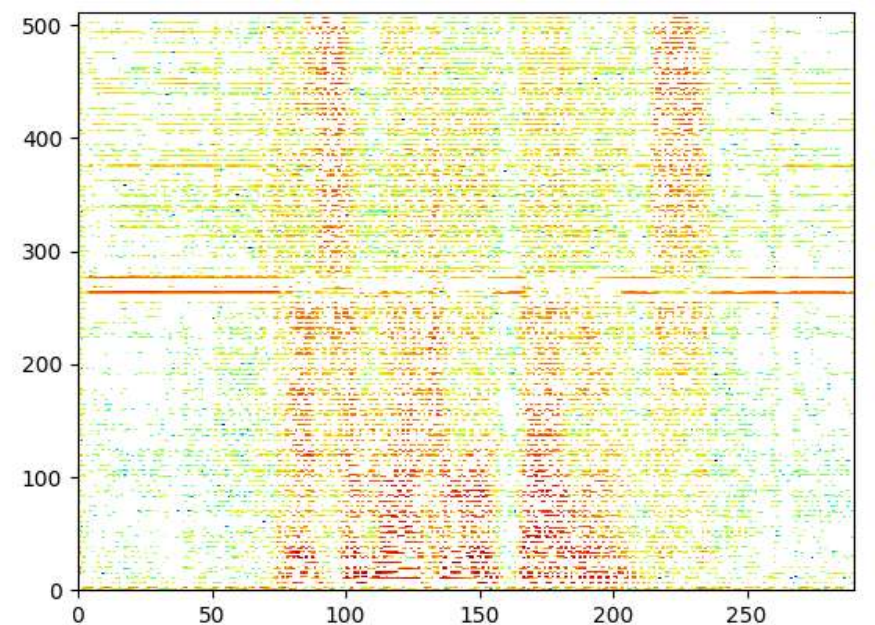}}
\caption{ Mask.$1_{th}$ (GFT-SVD-CNet)}
\end{subfigure}
\vskip 0.1in
\begin{subfigure}[t]{0.48\columnwidth}
\centerline{\includegraphics[width=\columnwidth]{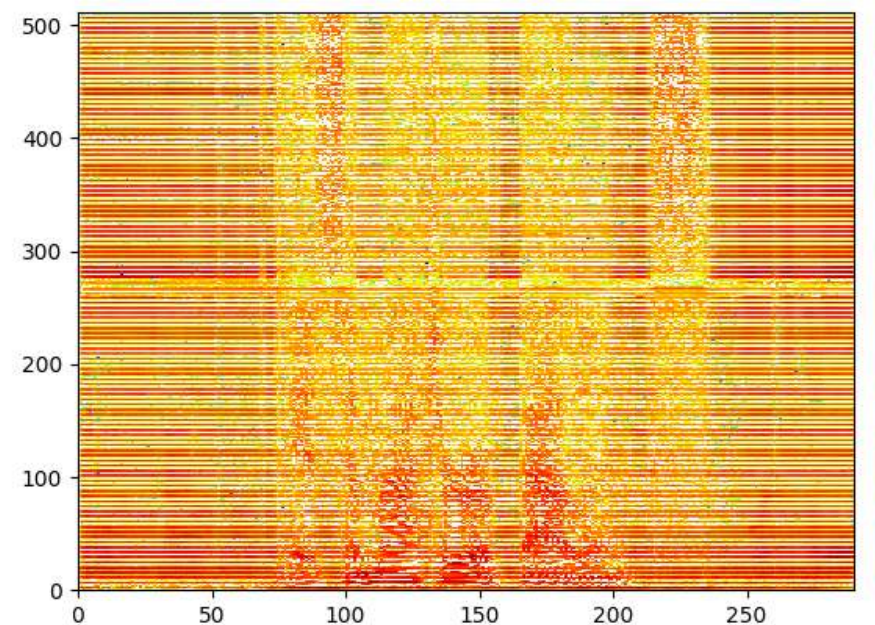}}
\caption{Mask.$1_{th}$ (GFT-SVD-CNet-LA)}
\end{subfigure}
\begin{subfigure}[t]{0.48\columnwidth}
\centerline{\includegraphics[width=\columnwidth]{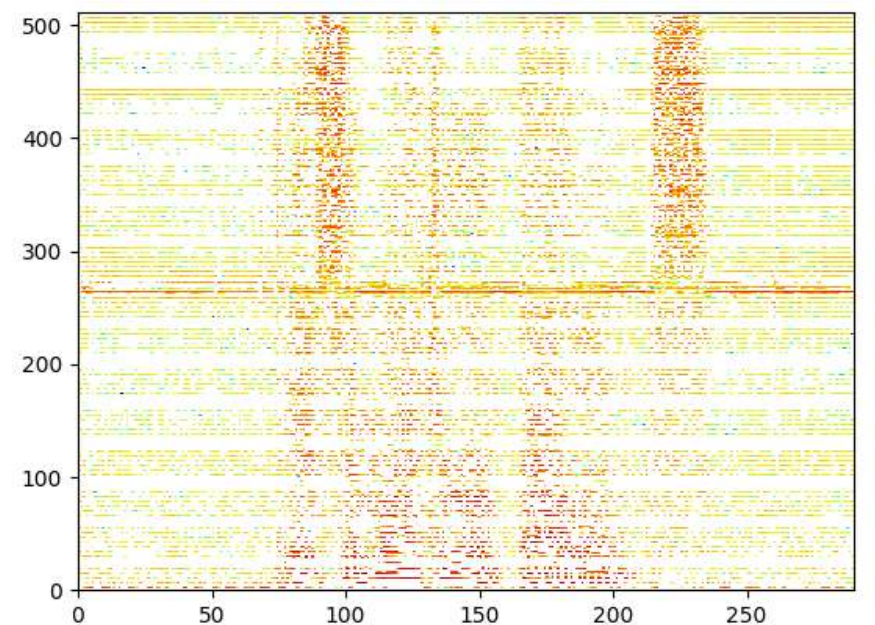}}
\caption{Mask.$2_{th}$ (GFT-SVD-CNet-LA)}
\end{subfigure}
\caption{\textcolor{black}{The illustrations of (a) the real and (b) imaginary parts of the estimated mask obtained by DPCRN using STFT. (c), (d) The estimated masks generated by DPCRN with GFT-SVD-LA. (e), (f) The estimated masks generated by DPCRN with GFT-SVD-CNet. (g), (h) The estimated masks generated by DPCRN with GFT-SVD-CNet-LA.}}
\label{fig:mask}
\end{figure}

\begin{figure}[!t]
\centering
\begin{subfigure}[t]{0.48\columnwidth}
\captionsetup{justification=centering}
\centerline{\includegraphics[width=\columnwidth]{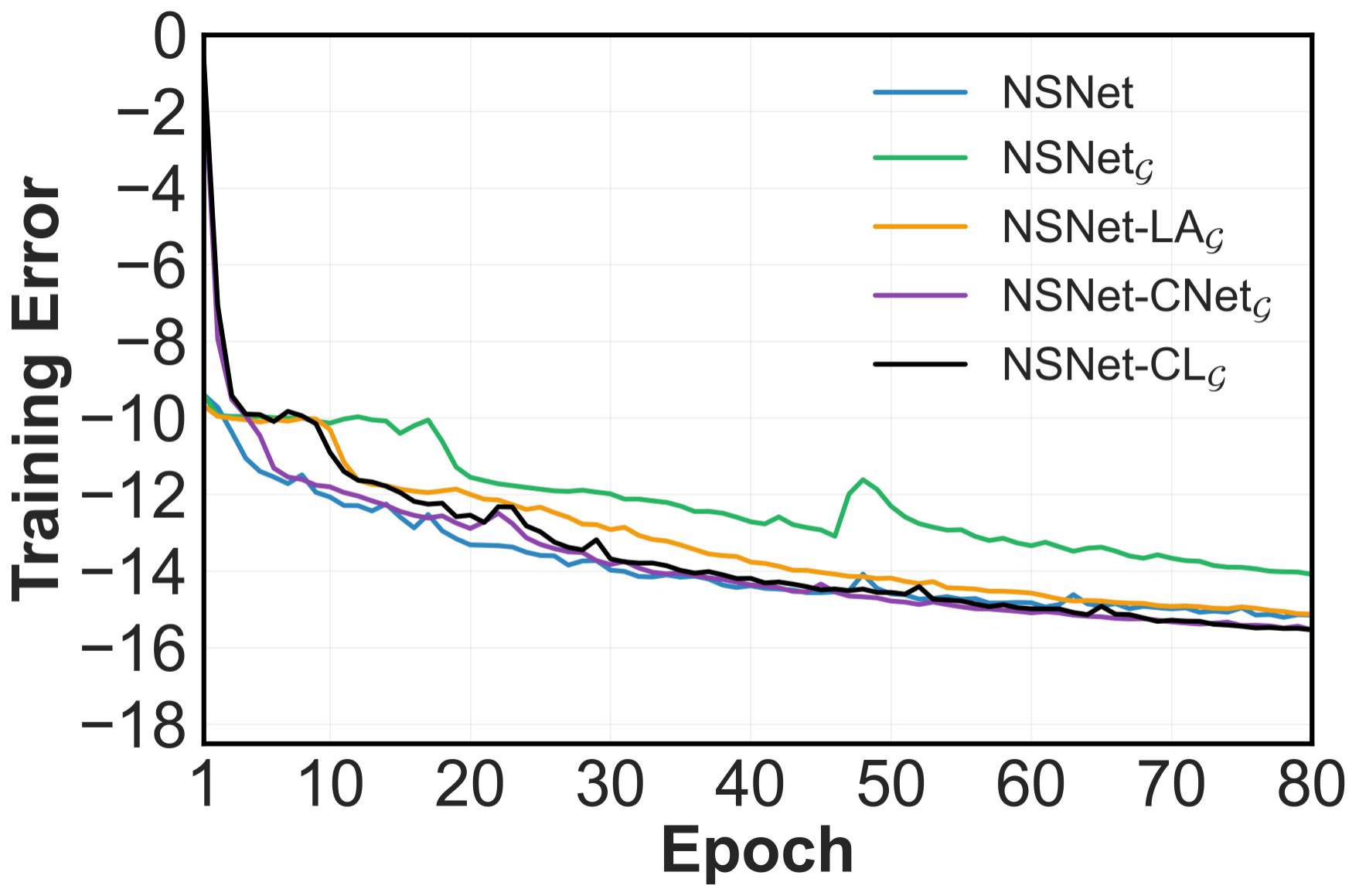}}
\caption{}
\end{subfigure}
\begin{subfigure}[t]{0.48\columnwidth}
\centerline{\includegraphics[width=\columnwidth]{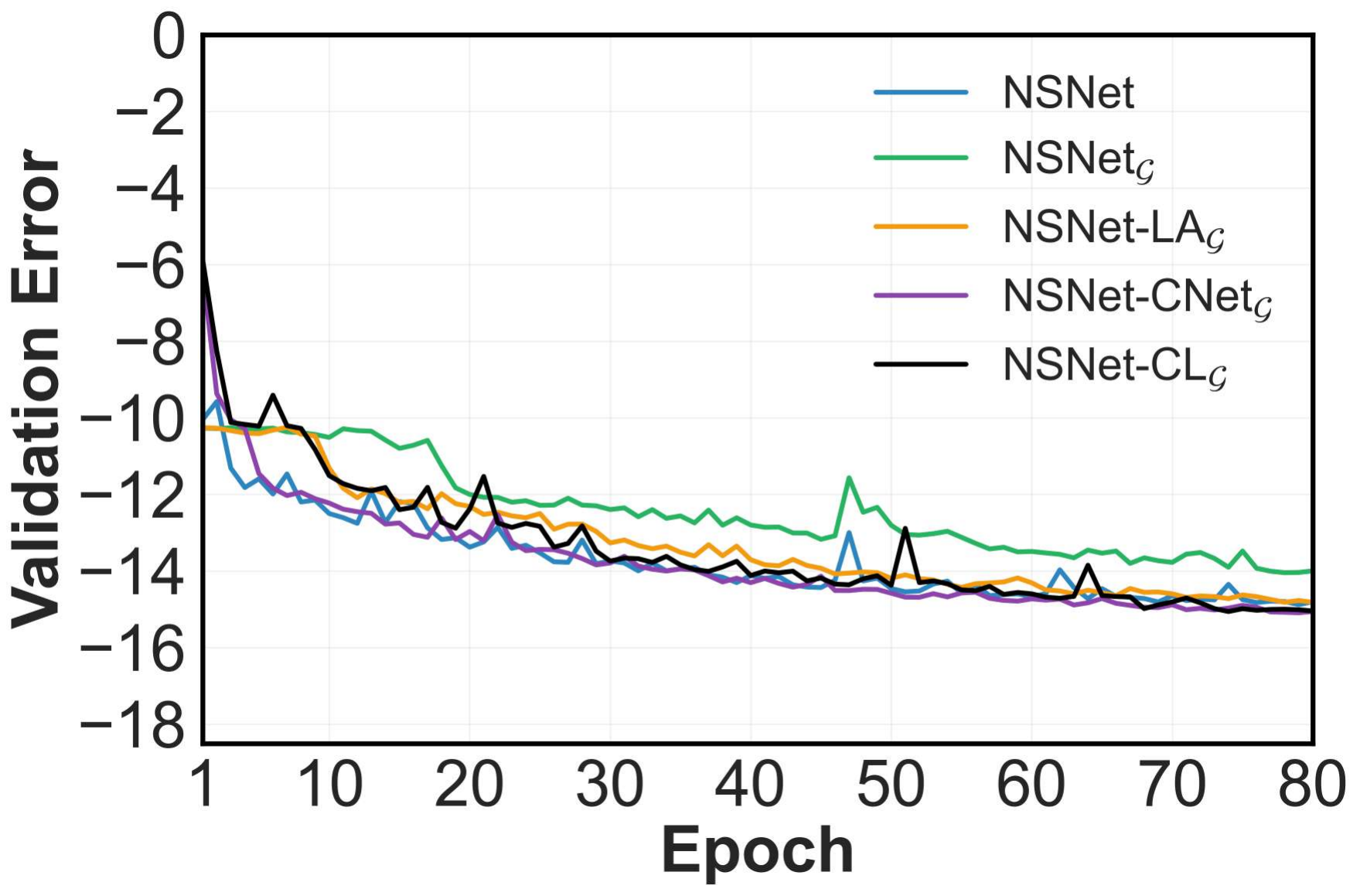}}
\caption{}
\end{subfigure}
\vskip 0.1in
\begin{subfigure}[t]{0.48\columnwidth}
\centerline{\includegraphics[width=\columnwidth]{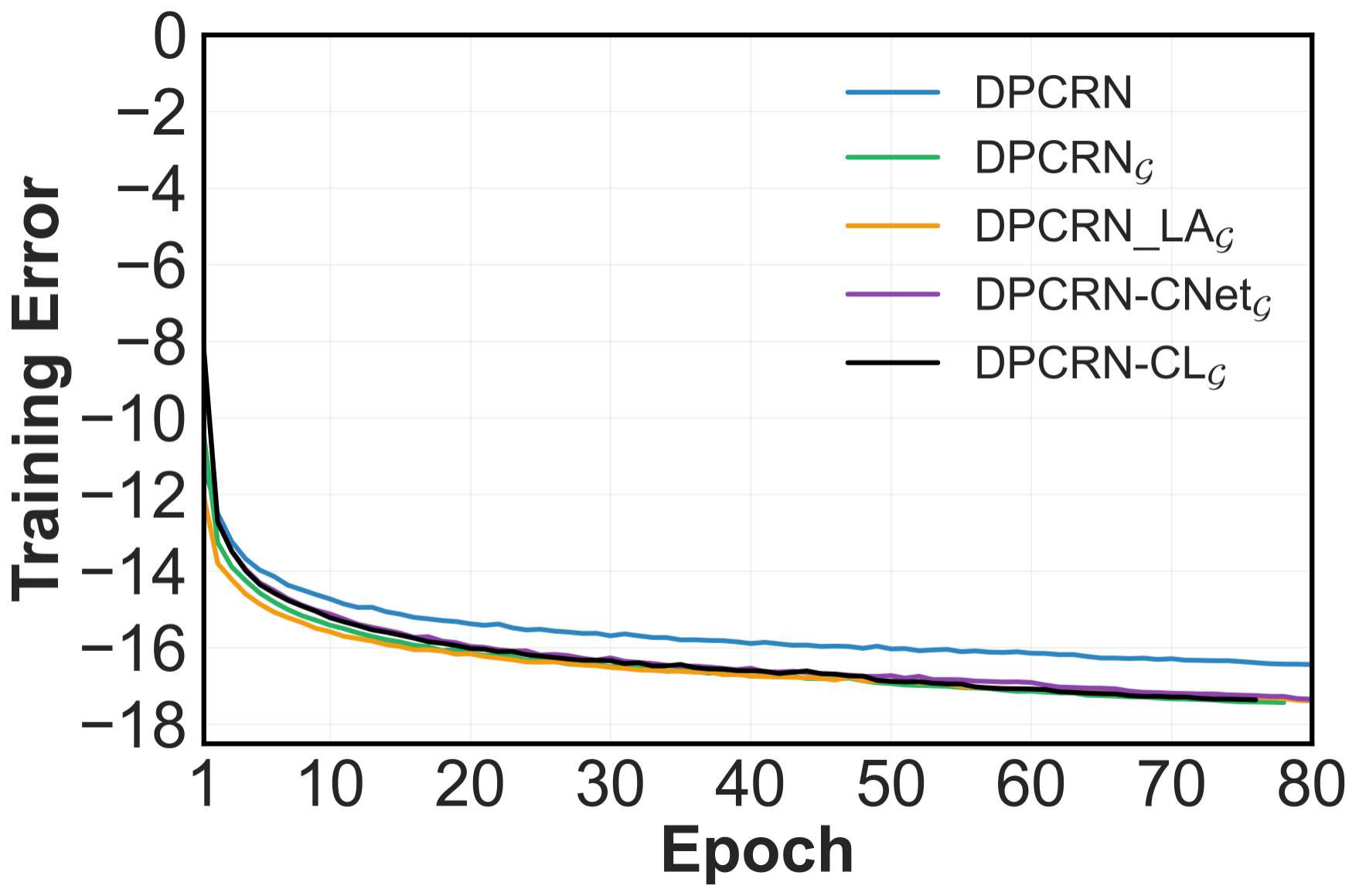}}
\caption{}
\end{subfigure}
\begin{subfigure}[t]{0.48\columnwidth}
\centerline{\includegraphics[width=\columnwidth]{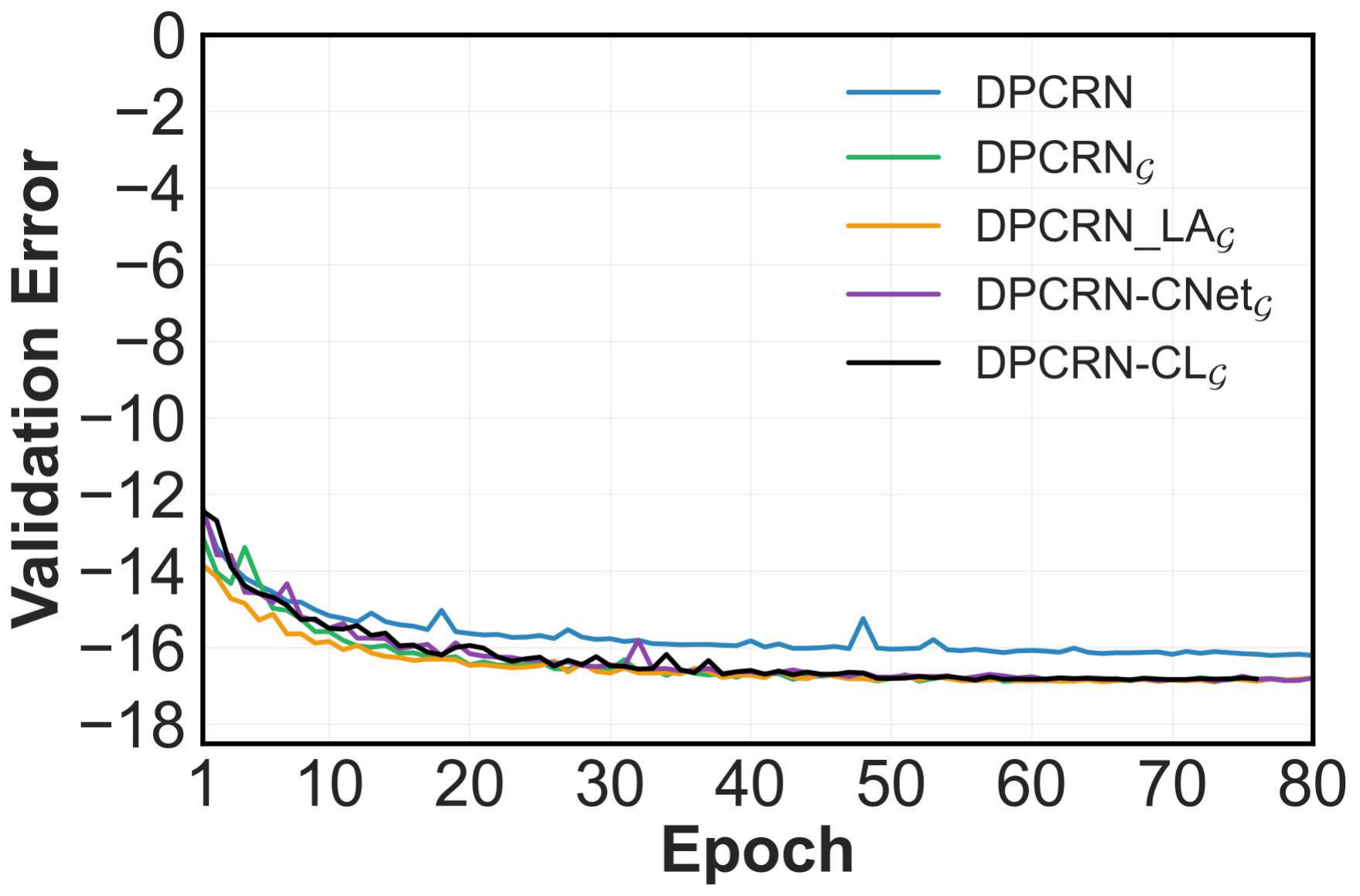}}
\caption{}
\end{subfigure}
\vskip 0.1in
\begin{subfigure}[t]{0.48\columnwidth}
\centerline{\includegraphics[width=\columnwidth]{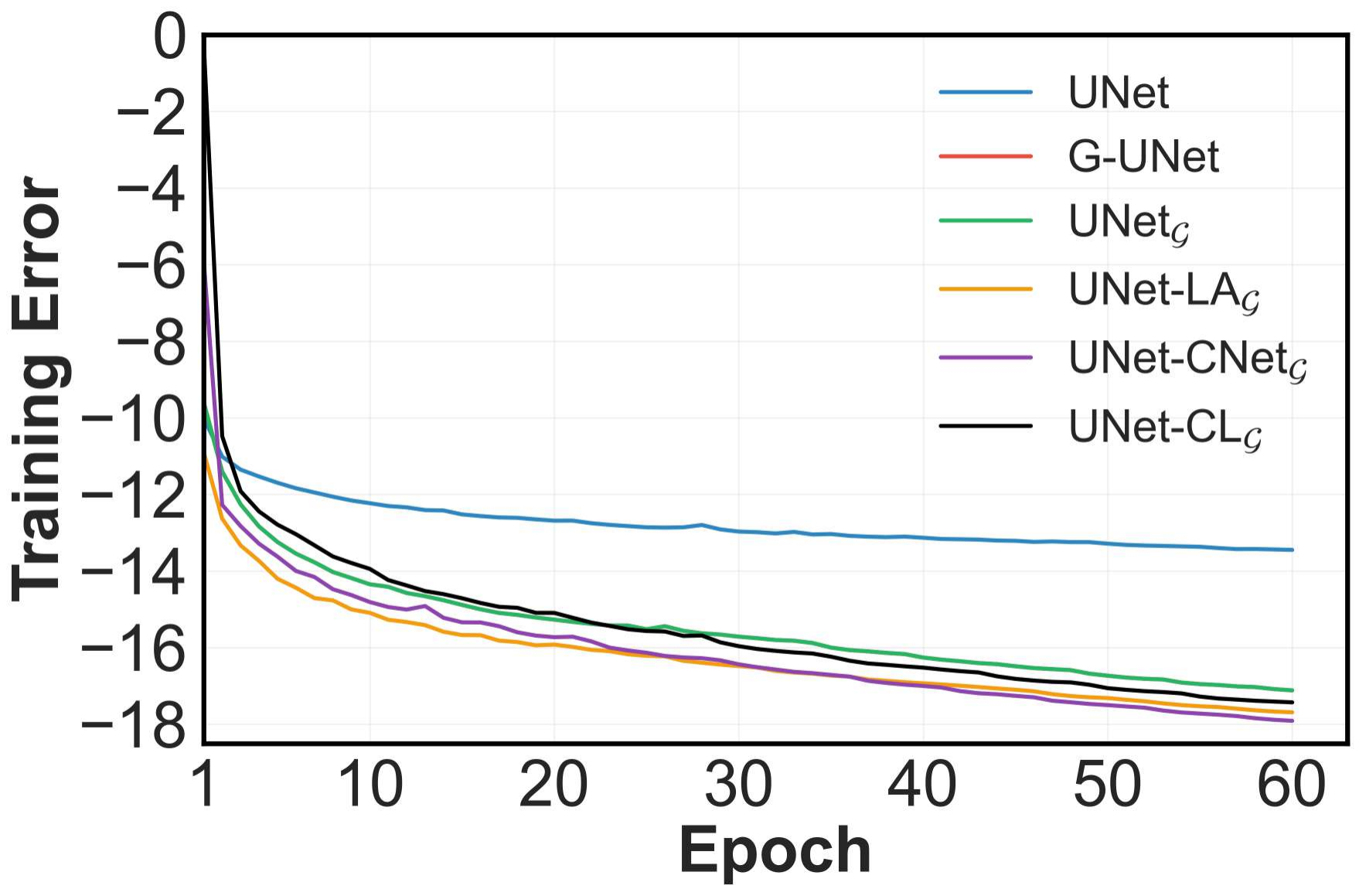}}
\caption{}
\end{subfigure}
\begin{subfigure}[t]{0.48\columnwidth}
\centerline{\includegraphics[width=\columnwidth]{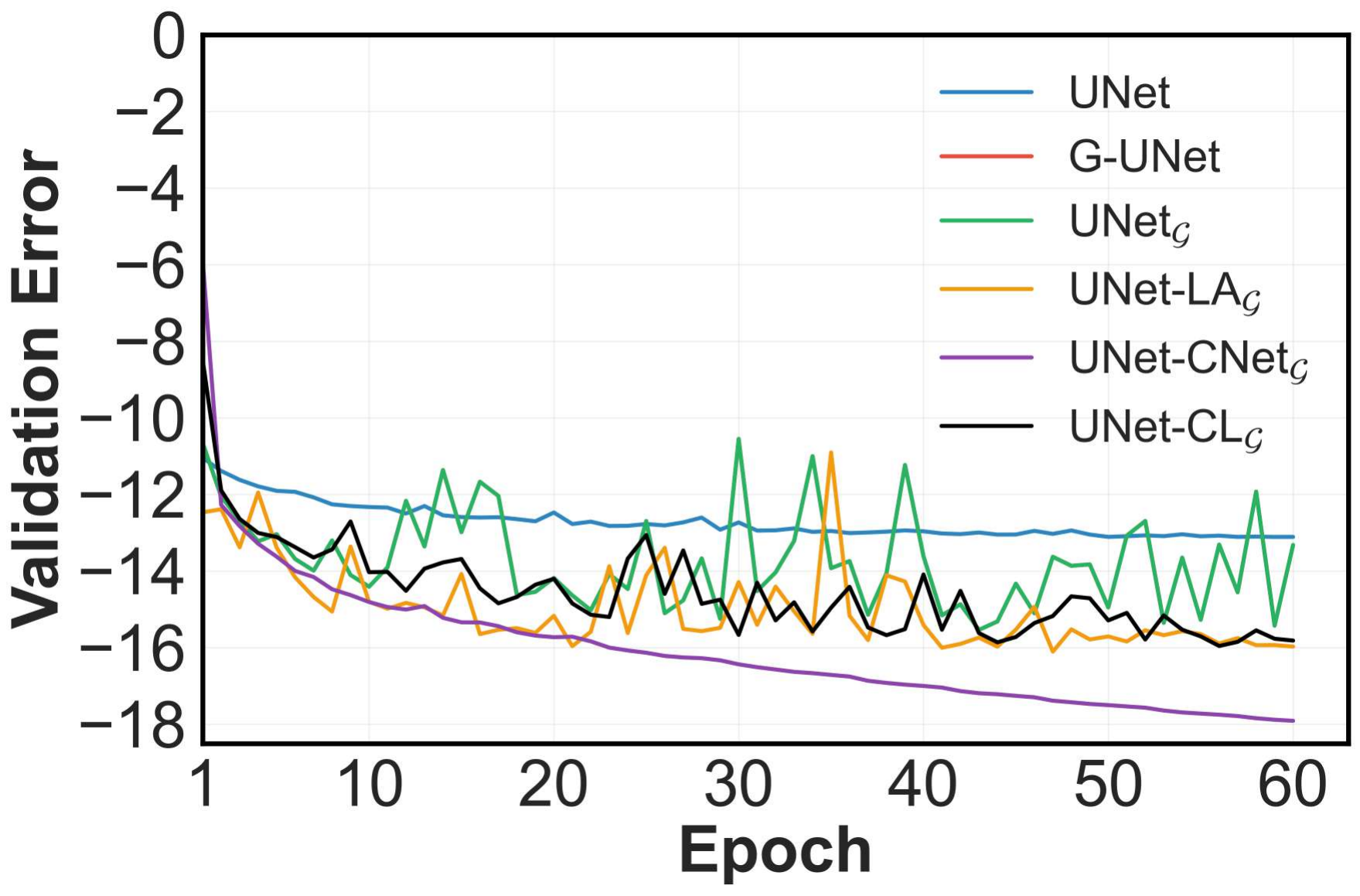}}
\caption{}
\end{subfigure}
\vskip 0.1in
\begin{subfigure}[t]{0.48\columnwidth}
\centerline{\includegraphics[width=\columnwidth]{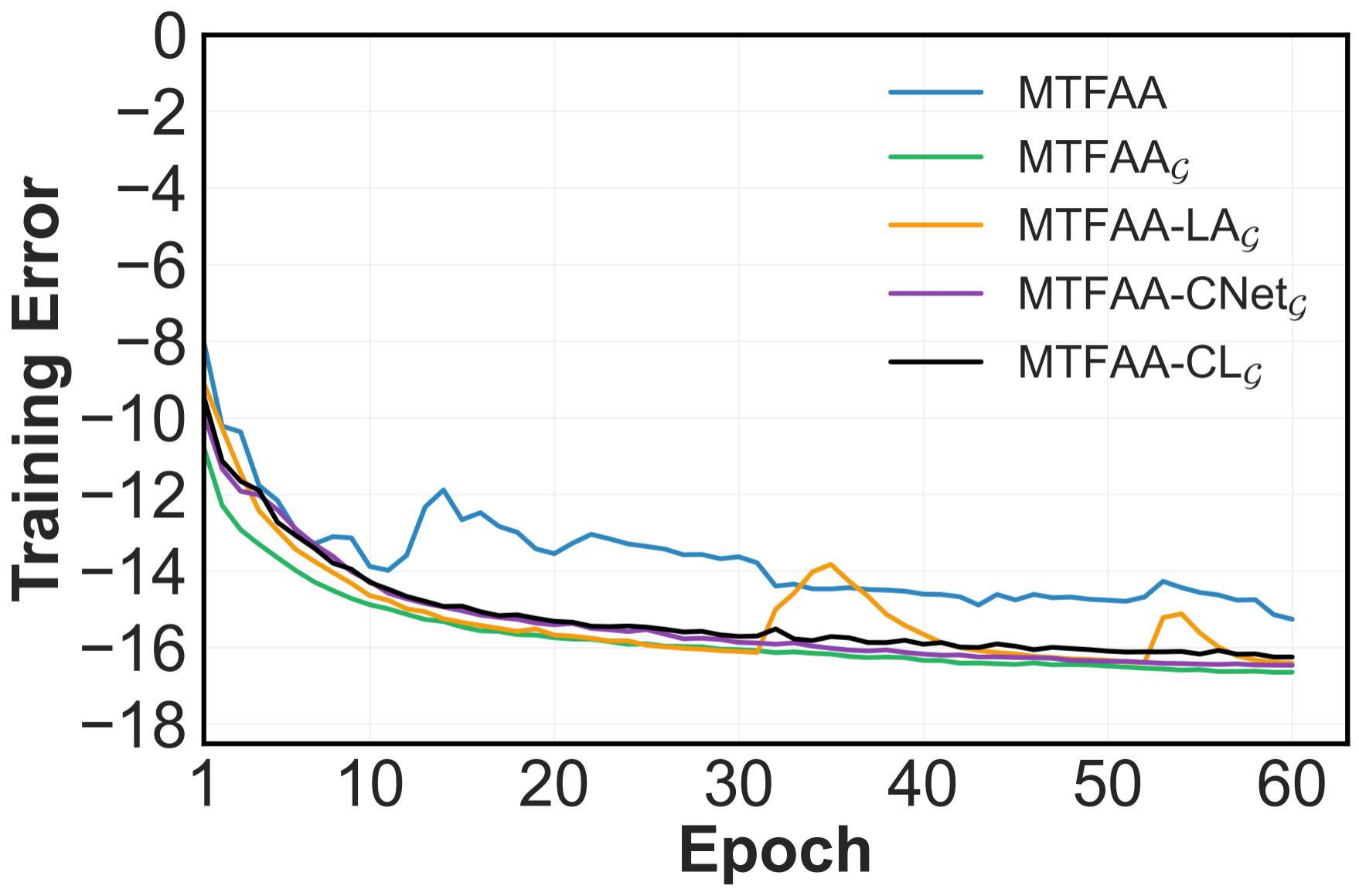}}
\caption{}
\end{subfigure}
\begin{subfigure}[t]{0.48\columnwidth}
\centerline{\includegraphics[width=\columnwidth]{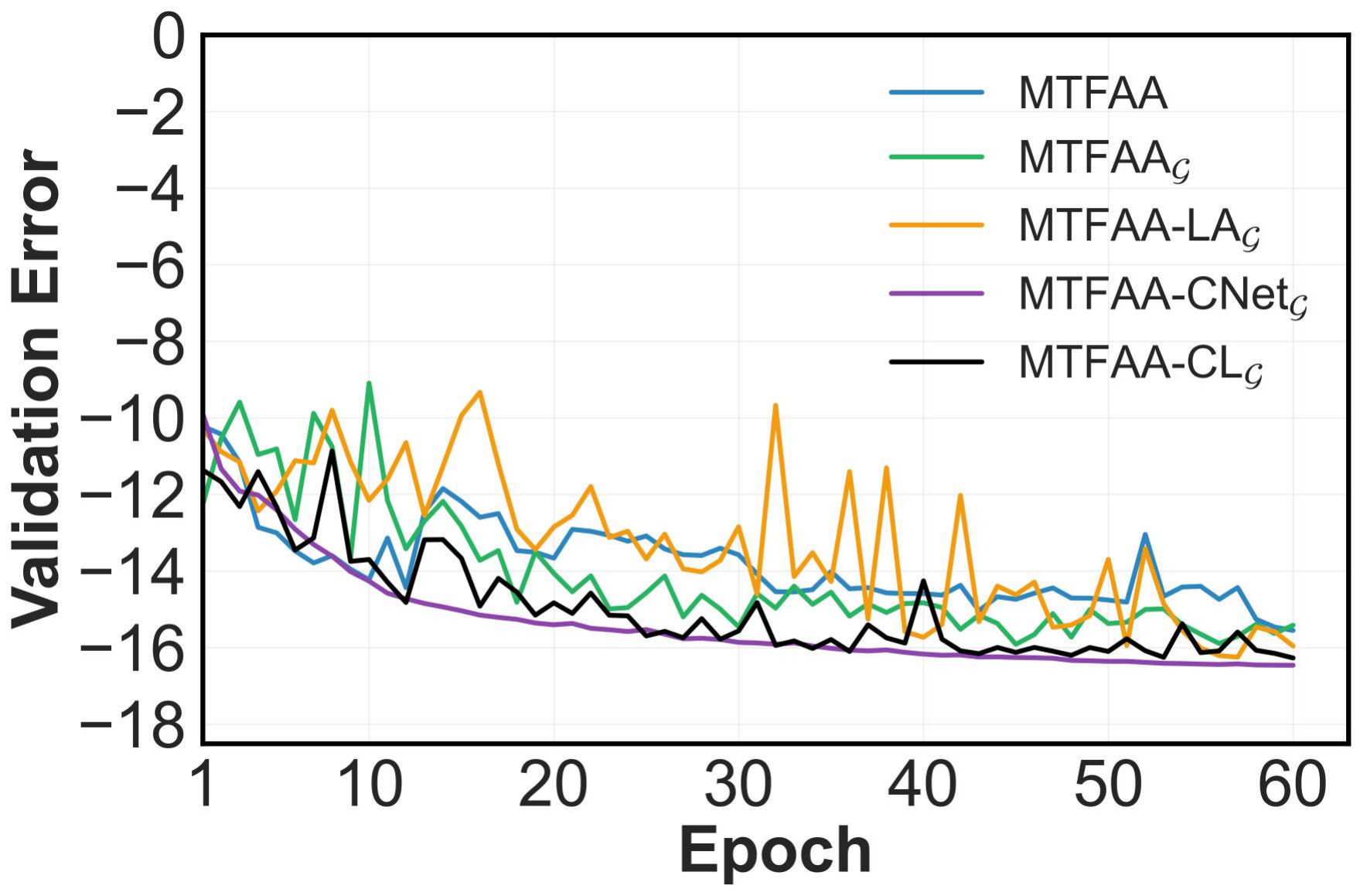}}
\caption{}
\end{subfigure}
\vskip 0.1in
\begin{subfigure}[t]{0.48\columnwidth}
\centerline{\includegraphics[width=\columnwidth]{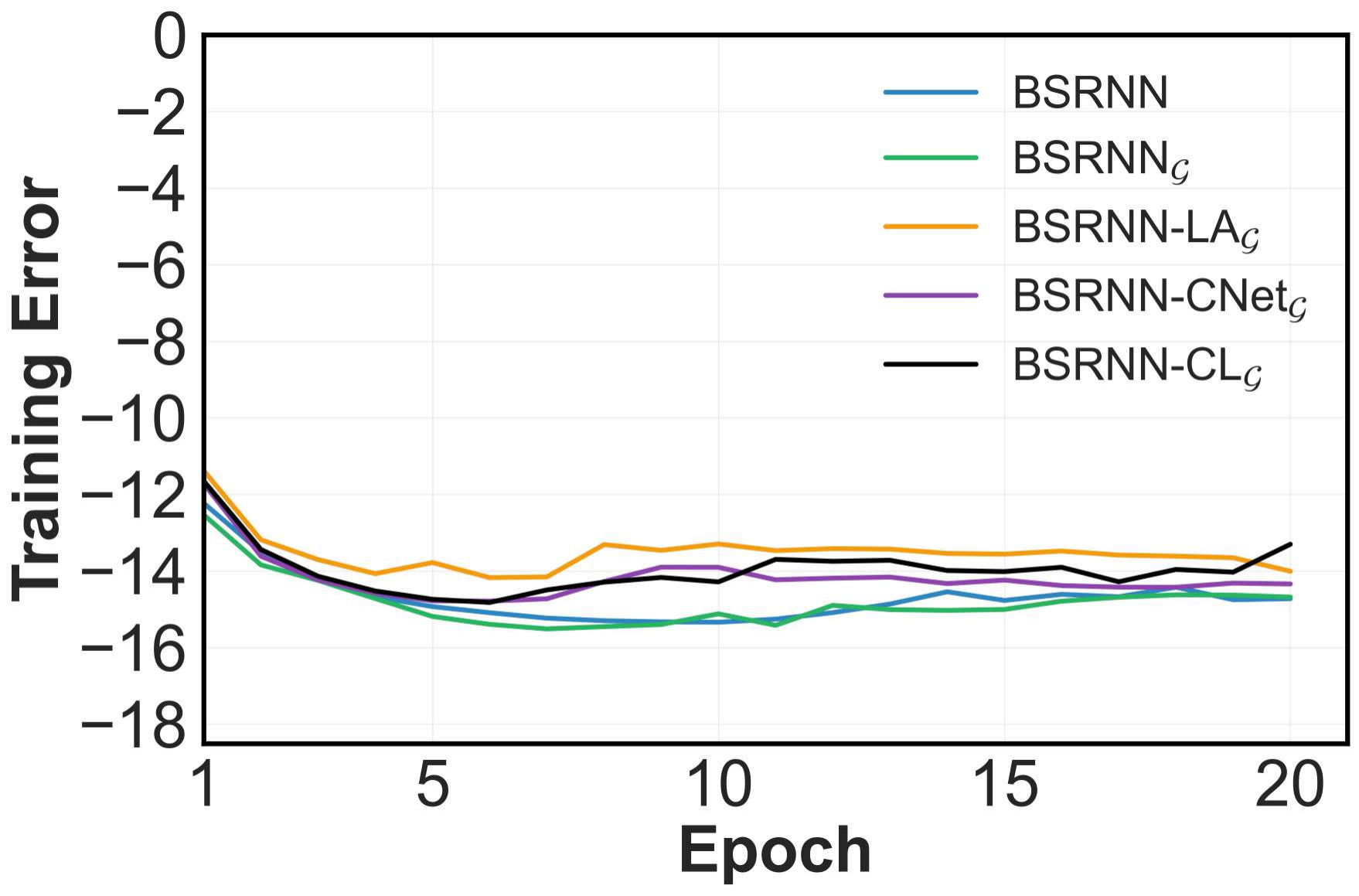}}
\caption{}
\end{subfigure}
\begin{subfigure}[t]{0.48\columnwidth}
\centerline{\includegraphics[width=\columnwidth]{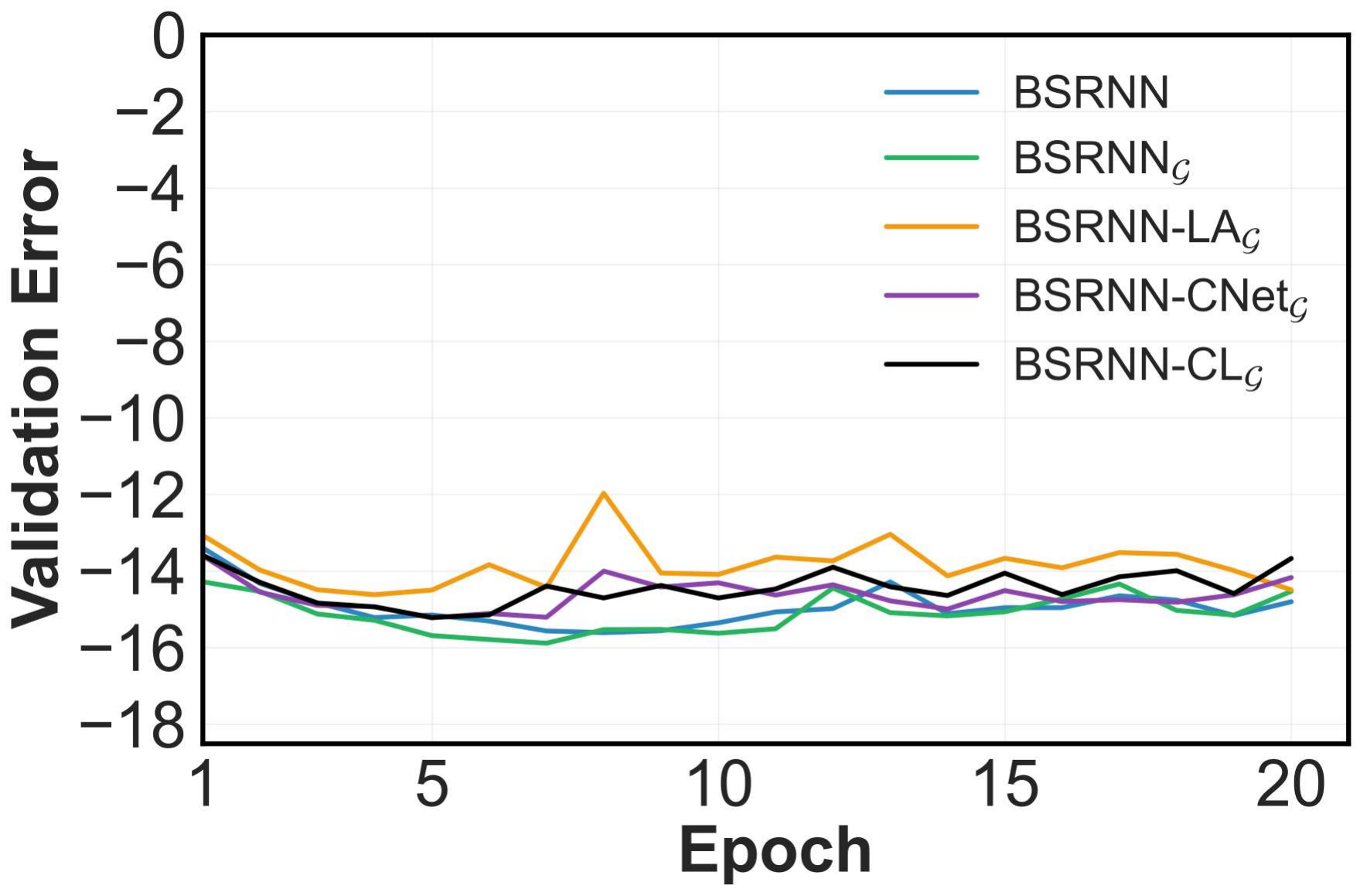}}
\caption{}
\end{subfigure}
\caption{\textcolor{black}{The curves of (a) training error and (b) validation error of NSNet using various Fourier transform. The curves of training error (c) validation error (d) on the combination of DPCRN and various Fourier transform. The curves of training error (e) validation error (f) on the combination of UNet and various Fourier transform. The curves of training error (g) validation error (h) on the combination of MTFAA and various Fourier transform. The curves of training error (i) validation error (j) on the combination of BSRNN and various Fourier transform.}}
\label{fig:loss}
\end{figure}

\textcolor{black}{\subsection{Visualization}
\textcolor{black}{In Fig.~\ref{fig:mask}, we present visualization of the masks estimated by the DPCRN backbone with various Fourier representations, where DPCRN separately predicts the magnitude mask and phase of the clean speech. 
\textcolor{black}{The noisy waveform used in Fig.~\ref{fig:mask} is generated by mixing a clean speech utterance (Fig.~\ref{fig:spectrm1}) with a noise recording at 5dB SNR.} Fig.~\ref{fig:mask} (a) and (b) illustrates the real and imaginary parts of the mask attained by DPCRN-STFT, respectively. Fig.~\ref{fig:mask} (c)-(h) show the estimated masks produced by DPCRN using GFT-SVD-LA, GFT-SVD-CNet and GFT-SVD-CNet-LA. It can be clearly found that the STFT-based mask estimation entails separate optimization space toward the magnitude and phase. In contrast, the DPCRN with learnable GFT-SVD is capable of estimating a unified mask for both magnitude and phase. This alleviates the compensation effects that decouple the complex spectrum estimation into two separate steps. Meanwhile, compared to the mask estimated by DPCRN with GFT-SVD-LA or GFT-SVD-CNet, the GFT-SVD-CNet-LA approach capture finer details in the learnable GFT domain. Therefore, the comparisons among STFT, GFT-SVD-LA, GFT-SVD-CNet, and GFT-SVD-CNet-LA demonstrate the efficacy of the learnable graph topology with a learnable graph Fourier basis.}}
\textcolor{black}{\subsection{Training \& Validation Error}
Fig.~\ref{fig:loss} shows the curves of training and validation errors produced by each backbone, \textcolor{black}{where each mini-batch includes 20 \textcolor{black}{samples}. From these curves, we can observe that backbone models with our proposed learnable GFT-SVD-CNet-LA (-CL$_G$) yields lower training and validation errors compared to the backbones using STFT, GFT-EVD, GFT-SVD, confirming the efficacy of the learnable GFT-SVD. Meanwhile, we can find that compared to learnable GFT-SVD-LA (-LA$_G$) and GFT-SVD-CNet (-CNet$_G$), GFT-SVD-CNet-LA yields lower training and validation errors.}}

\subsection{Computational Complexity}
In Table~\ref{table_par}, we report the multiplier-accumulator operations (MACs), the number of parameters, and the real-time factor (RTF) for different backbone networks with STFT, GFT-EVD, GFT-SVD, Conv-1D and learnable GFT-SVD. Specifically, we employ 3-second audio segments as the input to measure the MACs per second~\cite{timegraph} and RTF, with the latter measured on an NVIDIA GeForce GTX 3090 Ti. It can be observed that the use of learnable GFT-SVD does not substantially compromise the computational efficiency of backbone networks in MACs, model size, and RTF. Meanwhile, as the evaluation results reported in Tables~\ref{table_Voicebank+Demand} and~\ref{table_DNSchallengs}, the learnable GFT-SVD provides substantial performance improvements.

\section{Conclusion} \label{Conclusion}
\label{sec:conclusion}
The alignment performance of magnitude and phase in the GFT domain depends on the accuracy of the employed graph Fourier basis for GFT-based neural speech enhancement system. Motivated by this, we investigate a learnable graph topology initialized with graph shift operators to dynamically characterize relationships among speech samples. A learnable Conv-1D Fourier basis inversion is then proposed to reduce matrix inversion errors in GFT. Extensive experiments on the DNS-2020 and VCTK+DEMAND benchmarks demonstrate that state-of-the-art neural speech enhancement architectures with the learnable GFT-SVD outperform those using STFT, GFT-EVD, GFT-SVD, and Conv-1D \textcolor{black}{in three widely used objective evaluation metrics. This study provides a new direction for modeling the alignment of spectral magnitude and phase to better learn clean spectrogram information in neural speech enhancement.}

\begin{table}[!t]
	\caption{\textcolor{black}{The MACs, parameter numbers, and real-time factor (RTF) of different backbone networks with STFT, GFT-SVD, and learnable GFT-SVD.}}
     \centering
    \def\arraystretch{1.1}
    \setlength{\tabcolsep}{1.5pt}
  \begin{tabular}{lcccc}
    \toprule[1.0pt] 
   \bf{Network} & \bf Transform  & \bf{MACs} (G/s) & \#\bf{Param} (M)& \multirow{1}*{\bf{RTF}}\\
   \midrule \midrule
        NSNet~\cite{NSNet}  & \multirow{1}*{STFT} &{1.15}&{3.04}& {0.009}\\
         {NSNet$_{\cal {G}}$}~\cite{timegraph}&GFT-SVD &1.15&3.04& 0.006\\
          {NSNet-CL$_{\cal {G}}$}&GFT-SVD-CNet-LA&2.50&5.41&0.018 \\
         \midrule
		DPCRN~\cite{DPCRN}& STFT&{37.42}&{0.88}& {0.011}\\
        {DPCRN$_{\cal {G}}$}~\cite{timegraph}&GFT-SVD&37.62&0.88&0.005  \\
   {DPCRN-CL$_{\cal {G}}$}&GFT-SVD-CNet-LA&38.97&3.25&0.010  \\
  \midrule
		DCRN~\cite{DCRN}& STFT& {16.56}&{2.03}& {0.009} \\
        {DCRN$_{\cal {G}}$}~\cite{timegraph}&GFT-SVD&16.65&2.03&{0.003}  \\ 
  {DCRN-CL$_{\cal {G}}$}&GFT-SVD-CNet-LA&17.99&4.39&{0.030}  \\ 
  \midrule
		DCCRN~\cite{DCCRN}& STFT&{65.85}& {4.33}& {0.031}\\
  {DCCRN$_{\cal {G}}$}~\cite{timegraph}&GFT-SVD&66.19 &4.33 &0.013 \\
 {DCCRN-CL$_{\cal {G}}$}&GFT-SVD-CNet-LA&67.54 &6.69 &0.071 \\
  \midrule
	  MTFAA~\cite{MTFAA}& STFT& {14.55}&{2.19}& {0.040}\\
   	{MTFAA$_{\cal {G}}$}~\cite{timegraph}&GFT-SVD&14.63 & 2.19 &0.012 \\
   {MTFAA-CL$_{\cal {G}}$}&GFT-SVD-CNet-LA&15.97 & 4.55 &0.051 \\
    \midrule
        BSRNN~\cite{bsrnn}& STFT& {2850}&{200}& {0.247}  \\
        {BSRNN$_{\cal {G}}$}~\cite{timegraph}&GFT-SVD&{2865}&{200}& {0.118}  \\ 
          {BSRNN-CL$_{\cal {G}}$}&GFT-SVD-CNet-LA&{2866}&{202}& {0.551}  \\ 
         \midrule
    UNet~\cite{Unet} &STFT&22.52 &18.10&{0.019} \\ 
            G-UNet~\cite{ZhangP22}&GFT-EVD&22.28 & 18.10&0.025  \\     
            {UNet$_{\cal {G}}$}~\cite{timegraph} &GFT-SVD&19.38 &18.00&{0.009} \\  
             {UNet-CL$_{\cal {G}}$} &GFT-SVD-CNet-LA&23.98 &20.46&{0.018} \\ 
                \midrule
          \textcolor{black}{{{Conv-TasNet}}~\cite{LuoM19}}  &  \textcolor{black}{Conv-1D}& \textcolor{black}{33.54} & \textcolor{black}{3.33 }& \textcolor{black}{0.038} \\
          \textcolor{black}{{{Conv-TasNet}}} & \textcolor{black}{STFT} & \textcolor{black}{5.549} & \textcolor{black}{8.831 }& \textcolor{black}{0.045} \\
         \textcolor{black}{{{Conv-TasNet}$_{\cal {G}}$}} &\textcolor{black}{GFT-SVD}&{\textcolor{black}{5.566}} &{\textcolor{black}{8.831}}&{\textcolor{black}{0.042}}  \\
        \textcolor{black}{{{Conv-TasNet-CL}$_{\cal {G}}$}} &\textcolor{black}{GFT-SVD-CNet-LA} &{\textcolor{black}{6.915}} &{\textcolor{black}{11.195}}&{\textcolor{black}{0.041}}  \\ 
        \midrule
           \textcolor{black}{{{DPTNet}}~\cite{DPTNet}}  &\textcolor{black}{STFT} & \textcolor{black}{179.84} & \textcolor{black}{0.87}& \textcolor{black}{0.086} \\
         \textcolor{black}{{{DPTNet}$_{\cal {G}}$}} &\textcolor{black}{GFT-SVD}&{\textcolor{black}{180.42}} &{\textcolor{black}{0.87}}&{\textcolor{black}{0.039}}  \\
        \textcolor{black}{{{DPTNet}-CL$_{\cal {G}}$}} &\textcolor{black}{GFT-SVD-CNet-LA} &{\textcolor{black}{181.77}} &{\textcolor{black}{3.23}}&{\textcolor{black}{0.041}}  \\ 
      \toprule[1.0pt]
  \end{tabular}
  \label{table_par}
\end{table}
\bibliographystyle{IEEEtran}
\bibliography{refs1.bib}

\begin{thebibliography}{10}
\providecommand{\url}[1]{#1}
\csname url@samestyle\endcsname
\providecommand{\newblock}{\relax}
\providecommand{\bibinfo}[2]{#2}
\providecommand{\BIBentrySTDinterwordspacing}{\spaceskip=0pt\relax}
\providecommand{\BIBentryALTinterwordstretchfactor}{4}
\providecommand{\BIBentryALTinterwordspacing}{\spaceskip=\fontdimen2\font plus
\BIBentryALTinterwordstretchfactor\fontdimen3\font minus \fontdimen4\font\relax}
\providecommand{\BIBforeignlanguage}[2]{{%
\expandafter\ifx\csname l@#1\endcsname\relax
\typeout{** WARNING: IEEEtran.bst: No hyphenation pattern has been}%
\typeout{** loaded for the language `#1'. Using the pattern for}%
\typeout{** the default language instead.}%
\else
\language=\csname l@#1\endcsname
\fi
#2}}
\providecommand{\BIBdecl}{\relax}
\BIBdecl

\bibitem{loizou}
P.~C. Loizou, \emph{Speech Enhancement: Theory and Practice}, 2nd~ed.\hskip 1em plus 0.5em minus 0.4em\relax Boca Raton, FL, USA: CRC Press, Inc., 2013.

\bibitem{TimeFrequency}
Q.~Zhang, X.~Qian, Z.~Ni, A.~Nicolson, E.~Ambikairajah, and H.~Li, ``A time-frequency attention module for neural speech enhancement,'' \emph{IEEE/ACM Transactions on Audio, Speech, and Language Processing}, vol.~31, pp. 462--475, 2023.

\bibitem{R1}
A.~J. Fuglsig, J.~O. stergaard, J.~Jensen, L.~S. ndergaard Bertelsen, P.~Mariager, and Z.~H. Tan, ``Joint far- and near-end speech intelligibility enhancement based on the approximated speech intelligibility index,'' in \emph{Proc. IEEE Int. Conf. Acoust., Speech, Signal Process.}, 2022, pp. 7752--7756.

\bibitem{R3}
A.~J. Fuglsig, J.~O. stergaard, J.~Jensen, L.~S. ondergaard Bertelsen, P.~Mariager, and Z.~H. Tan, ``Joint far-and near-end speech intelligibility enhancement based on the approximated speech intelligibility index,'' \emph{CoRR}, vol. abs/2111.07759, 2021.

\bibitem{R4}
A.~J. Fuglsig, J.~Jensen, Z.~H. Tan, L.~S. ndergaard Bertelsen, J.~C. Lindof, and J.~stergaard, ``Joint far- and near-end speech and listening enhancement with minimum processing,'' in \emph{IEEE Access}, vol.~12, 2024, pp. 119\,983--120\,004.

\bibitem{R12}
S.~Shoba, R.~Rajavel, K.~Asutosh, and B.~Majhi, ``{Monaural speech separation using GA-DNN integration scheme},'' vol. 160, p. 107140, 2020.

\bibitem{mambadc}
M.~Chen, Q.~Zhang, M.~Wang, X.~Zhang, H.~Liu, E.~Ambikairaiah, and D.~Chen, ``{Selective State Space Model for Monaural Speech Enhancement},'' \emph{IEEE Trans. Consum. Electron.}, pp. 1--1, 2025.

\bibitem{mmse2017}
Krawczyk-Becker, Martin, and T.~Gerkmann, ``On mmse-based estimation of amplitude and complex speech spectral coefficients under phase-uncertainty,'' \emph{IEEE/ACM Trans. Audio, Speech, and Lang. Proc.}, vol.~24, no.~12, pp. 2251--2262, 2016.

\bibitem{mmse}
Y.~Ephraim and D.~Malah, ``{Speech Enhancement Using a Minimum Mean-Square Error Short-Time Spectral Amplitude Estimator},'' \emph{IEEE Trans. Acoust., Speech, Signal Process.}, vol. ASSP-32, no.~6, pp. 1109--1121, 1984.

\bibitem{zhang2019}
Q.~Zhang, M.~Wang, Y.~Lu, L.~Zhang, and M.~Idrees, ``A novel fast nonstationary noise tracking approach based on mmse spectral power estimator,'' \emph{Digital Signal Processing}, vol.~88, pp. 41--52, 2019.

\bibitem{zhang2019fast}
Q.~Zhang, M.~Wang, Y.~Lu, M.~Idrees, and L.~Zhang, ``Fast nonstationary noise tracking based on log-spectral power mmse estimator and temporal recursive averaging,'' \emph{IEEE Access}, vol.~7, pp. 80\,985--80\,999, 2019.

\bibitem{overview2018}
D.~Wang and J.~Chen, ``Supervised speech separation based on deep learning: An overview,'' \emph{IEEE/ACM Transactions on Audio, Speech, and Language Processing}, vol.~26, no.~10, pp. 1702--1726, 2018.

\bibitem{SpeechEnhancement}
O'Shaughnessy and Douglas, ``Speech enhancement—a review of modern methods,'' \emph{IEEE Transactions on Human-Machine Systems}, vol.~54, no.~1, pp. 110--120, 2024.

\bibitem{DeepMMSE}
Q.~Zhang, A.~Nicolson, M.~Wang, K.~K. Paliwal, and C.~Wang, ``{DeepMMSE}: A deep learning approach to mmse-based noise power spectral density estimation,'' \emph{IEEE/ACM Transactions on Audio, Speech, and Language Processing}, vol.~28, pp. 1404--1415, 2020.

\bibitem{chen2023neural}
M.~Chen, Q.~Zhang, Q.~Song, X.~Qian, R.~Guo, M.~Wang, and D.~Chen, ``Neural-free attention for monaural speech enhancement toward voice user interface for consumer electronics,'' \emph{IEEE Transactions on Consumer Electronics}, vol.~69, no.~4, pp. 765--774, 2023.

\bibitem{mamba}
X.~Zhang, Q.~Zhang, H.~Liu, T.~Xiao, X.~Qian, B.~Ahmed, E.~Ambikairajah, H.~Li, and J.~Epps, ``{Mamba in Speech: Towards an Alternative to Self-Attention},'' \emph{IEEE TASLP}, vol.~33, pp. 1933--1948, 2025.

\bibitem{XuDDL15}
Y.~Xu, J.~Du, L.~Dai, and C.~Lee, ``A regression approach to speech enhancement based on deep neural networks,'' \emph{IEEE/ACM Trans. Audio Speech Lang. Process.}, vol.~23, no.~1, pp. 7--19, 2015.

\bibitem{R5}
A.~Kar, S.~Sivapatham, and H.~Reddy, ``Improved monaural speech enhancement via low-complexity fully connected neural networks: A performance analysis,'' \emph{Circuits Syst. Signal Process.}, vol.~44, no.~5, pp. 3258--3287, 2025.

\bibitem{R9}
H.~Reddy, A.~Kar, and J.~Østergaard, ``Performance analysis of low complexity fully connected neural networks for monaural speech enhancement,'' \emph{Applied Acoustics}, vol. 190, p. 108627, 2022.

\bibitem{WangNW14}
Y.~Wang, A.~Narayanan, and D.~Wang, ``On training targets for supervised speech separation,'' \emph{IEEE/ACM Trans. Audio Speech Lang. Process.}, vol.~22, no.~12, pp. 1849--1858, 2014.

\bibitem{R6}
S.~Balasubramanian, R.~Rajavel, and A.~Kar, ``Ideal ratio mask estimation based on cochleagram for audio-visual monaural speech enhancement,'' \emph{Applied Acoustics}, vol. 211, p. 109524, 2023.

\bibitem{R7}
------, ``Estimation of ideal binary mask for audio-visual monaural speech enhancement,'' \emph{Circuits Syst. Signal Process.}, vol.~42, no.~9, pp. 5313--5337, 2023.

\bibitem{Performance}
S.~Sivapatham, A.~Kar, and R.~Ramadoss, ``Performance analysis of various training targets for improving speech quality and intelligibility,'' \emph{Applied Acoustics}, vol. 175, p. 107817, 2021.

\bibitem{PaliwalWS11}
K.~Paliwal, K.~Wójcicki, and B.~S. 1, ``The importance of phase in speech enhancement,'' \emph{Speech Commun.}, vol.~53, no.~4, pp. 465--494, 2011.

\bibitem{8466892}
N.~Zheng and X.-L. Zhang, ``{Phase-Aware Speech Enhancement Based on Deep Neural Networks},'' \emph{IEEE/ACM Transactions on Audio, Speech, and Language Processing}, vol.~27, no.~1, pp. 63--76, 2019.

\bibitem{R10}
S.~Devi, S.~Shoba, K.~Asutosh, and M.~Vladimir, ``Mask estimation using phase information and inter-channel correlation for speech enhancement,'' \emph{Circuits Systems and Signal Processing}, vol.~41, no.~7, pp. 4117--4135, 2022.

\bibitem{R11}
S.~Shoba, K.~Asutosh, B.~Roshan, M.~Vladimir, and S.~Pitikhate, ``A deep neural network-correlation phase sensitive mask based estimation to improve speech intelligibility,'' vol. 212, p. 109592, 2023.

\bibitem{SEGAN}
S.~Pascual, A.~Bonafonte, and J.~Serr`a, ``Segan: Speech enhance ment generative adversarial network,'' in \emph{in Proc. INTERSPEECH}, 2017, p. 3642–3646.

\bibitem{Endtoend}
S.-W. Fu, T.-W. Wang, Y.~Tsao, X.~Lu, and H.~Kawai, ``End-to-end wave form utterance enhancement for direct evaluation metrics optimization by fully convolutional neural networks,'' \emph{IEEE/ACM Trans. Audio, Speech, Lang. Process.}, vol.~26, p. 1570–1584, 2018.

\bibitem{Onlossfunctions}
M.~Kolbk, Z.~H. Tan, S.~H. Jensen, and J.~Jensen, ``On loss functions for supervised monaural time-domain speech enhancement,'' \emph{IEEE/ACM Trans. Audio, Speech, Lang. Process.}, vol.~28, pp. 825--838, 2020.

\bibitem{TCNN}
A.~Pandey and D.~Wang, ``Tcnn:temporalconvolutional neural network for real-time speech enhancement in the time domain,'' in \emph{in Proc. IEEE Int. Conf. Acoust., Speech Signal Process.}, 2019, p. 6875–6879.

\bibitem{LuoM19}
Y.~Luo and N.~Mesgarani, ``Conv-tasnet: Surpassing ideal time-frequency magnitude masking for speech separation,'' \emph{{IEEE} {ACM} Trans. Audio Speech Lang. Process.}, vol.~27, no.~8, pp. 1256--1266, 2019.

\bibitem{psm}
E.~Hakan, H.~J. R, W.~Shinji, and L.~R. Jonathan, ``Phase-sensitive and recognition-boosted speech separation using deep recurrent neural networks,'' in \emph{Proc. ICASSP}, 2015, pp. 708--712.

\bibitem{CRM}
D.~S. Williamson, Y.~Wang, and D.~Wang, ``Complex ratio masking for monaural speech separation,'' \emph{IEEE/ACM transactions on audio, speech, and language processing}, vol.~24, no.~3, pp. 483--492, 2015.

\bibitem{PHASEN}
D.~Yin, C.~Luo, Z.~Xiong, and W.~Zeng, ``Phasen:a phase-and-harmonics-aware speech enhancement network,'' in \emph{Proc. The Thirty-Fourth Conference on Artificial Intelligence (AAAI)}, 2020, pp. 9458--9465.

\bibitem{MPSENet}
Y.-X. Lu, A.~Yang, and Z.-H. Ling, ``Mp-senet: A speech enhancement model with parallel denoising of magnitude and phase spectra,'' in \emph{Proc. INTERSPEECH 2023}, 2023, pp. 3834--3838.

\bibitem{csm}
K.~Tan and D.~Wang, ``Learning complex spectral mapping with gated convolutional recurrent networks for monaural speech enhancement,'' \emph{{IEEE} {ACM} Trans. Audio Speech Lang. Process.}, vol.~28, pp. 380--390, 2020.

\bibitem{bsrnn}
J.~Yu, H.~Chen, Y.~Luo, R.~Gu, and C.~Weng, ``{High Fidelity Speech Enhancement with Band-split RNN},'' in \emph{Proc. INTERSPEECH}, 2023.

\bibitem{DCCRN}
Y.~Hu, Y.~Liu, S.~Lv, M.~Xing, S.~Zhang, Y.~Fu, J.~Wu, B.~Zhang, and L.~Xie, ``{DCCRN: Deep Complex Convolution Recurrent Network for Phase-Aware Speech Enhancement},'' in \emph{Proc. INTERSPEECH}, 2020, pp. 2472--2476.

\bibitem{WangWR21}
Z.~Q. Wang, G.~Wichern, and J.~L. Roux, ``The compensation between magnitude and phase in speech separation,'' \emph{{IEEE} Signal Process. Lett.}, vol.~28, pp. 2018--2022, 2021.

\bibitem{Two-Stage}
A.~Li, C.~Zheng, R.~Peng, and X.~Li, ``Two heads are better than one: {A} two-stage approach for monaural noise reduction in the complex domain,'' \emph{CoRR}, vol. abs/2011.01561, 2020.

\bibitem{LiYYZL22}
A.~Li, S.~You, G.~Yu, C.~Zheng, and X.~Li, ``Taylor, can you hear me now? {A} taylor-unfolding framework for monaural speech enhancement,'' in \emph{Proceedings of the Thirty-First International Joint Conference on Artificial Intelligence}, 2022, pp. 4193--4200.

\bibitem{DBTNet}
G.~Yu, A.~Li, H.~Wang, Y.~Wang, Y.~Ke, and C.~Zheng, ``{DBT-Net: Dual-Branch Federative Magnitude and Phase Estimation With Attention-in-Attention Transformer for Monaural Speech Enhancement},'' \emph{IEEE/ACM Trans. Audio Speech Lang. Process.}, vol.~30, pp. 2629--2644, 2022.

\bibitem{ref133}
O.~Antonio, F.~Pascal, K.~Jelena, M.~J. MF, and V.~Pierre, ``Graph signal processing: overview, challenges, and applications,'' \emph{Proceedings of the IEEE}, vol. 106, no.~5, pp. 808--828, 2018.

\bibitem{xu22_odyssey}
L.~Xu, M.~Tian, X.~Guo, Z.~Shan, J.~Jia, Y.~Peng, J.~Yang, and R.~K. Das, ``{A novel feature based on graph signal processing for detection of physical access attacks},'' in \emph{Proc. The Speaker and Language Recognition Workshop}, 2022, pp. 107--111.

\bibitem{YanYWG20}
X.~Yan, Z.~Yang, T.~Wang, and H.~Guo, ``An iterative graph spectral subtraction method for speech enhancement,'' \emph{Speech Commun.}, vol. 123, pp. 35--42, 2020.

\bibitem{ZhangP22}
C.~Zhang and X.~Pan, ``Single-channel speech enhancement using graph fourier transform,'' in \emph{Proc. INTERSPEECH}, 2022, pp. 946--950.

\bibitem{timegraph}
T.~Wang, T.~Wang, M.~Ge, Q.~Zhang, Z.~Ge, and Z.~Yang, ``Time-graph frequency representation with singular value decomposition for neural speech enhancement,'' in \emph{Proc. IEEE Int. Conf. Acoust., Speech, Signal Process.}, 2025, pp. 1--5.

\bibitem{newmultilayer}
T.~Wang, H.~Guo, Q.~Zhang, and Z.~Yang, ``A new multilayer graph model for speech signals with graph learning,'' \emph{Digit. Signal Process.}, vol. 122, p. 103360, 2022.

\bibitem{WangGGZY23}
T.~Wang, H.~Guo, Z.~Ge, Q.~Zhang, and Z.~Yang, ``An {MMSE} graph spectral magnitude estimator for speech signals residing on an undirected multiple graph,'' \emph{{EURASIP} J. Audio Speech Music. Process.}, vol. 2023, no.~1, p.~7, 2023.

\bibitem{ref8}
T.~Wang, H.~Guo, B.~Lyv, and Z.~Yang, ``Speech signal processing on graphs: graph topology, graph frequency analysis and denoising,'' \emph{Chinese Journal of Electronics}, vol.~29, no.~1, pp. 1--11, 2020.

\bibitem{ref9}
T.~Wang, H.~Guo, X.~Yan, and Z.~Yang, ``Speech signal processing on graphs: the graph frequency analysis and an improved graph wiener filtering method,'' \emph{Speech Commun.}, vol. 127, pp. 82--91, 2021.

\bibitem{ref1}
A.~Ortega, P.~Frossard, J.~Kovacevic, J.~M.~F. Moura, and P.~Vandergheynst, ``Graph signal processing: overview, challenges, and applications,'' \emph{Proc. {IEEE}}, vol. 106, no.~5, pp. 808--828, 2018.

\bibitem{ZhangYKCWWE22}
Z.~Zhang, T.~Yoshioka, N.~Kanda, Z.~Chen, X.~Wang, D.~Wang, and S.~E. Eskimez, ``All-neural beamformer for continuous speech separation,'' in \emph{in Proc. IEEE Int. Conf. Acoust., Speech, Signal Process.}\hskip 1em plus 0.5em minus 0.4em\relax {IEEE}, 2022, pp. 6032--6036.

\bibitem{DNS2020}
C.~K.~A. Reddy, V.~Gopal, R.~Cutler, E.~Beyrami, R.~Cheng, H.~Dubey, S.~Matusevych, R.~Aichner, A.~Aazami, S.~Braun, P.~Rana, S.~Srinivasan, and J.~Gehrke, ``The {INTERSPEECH} 2020 deep noise suppression challenge: Datasets, subjective testing framework, and challenge results,'' in \emph{Proc. INTERSPEECH}, 2020, pp. 2492--2496.

\bibitem{voicebank}
C.~V. Botinhao, X.~Wang, S.~Takaki, and J.~Yamagishi, ``Investigating rnn-based speech enhancement methods for noise-robust text-to-speech,'' in \emph{in Proc. SSW, 2016}, 2016, pp. 146--152.

\bibitem{LvHZX21}
S.~Lv, Y.~Hu, S.~Zhang, and L.~Xie, ``{DCCRN+:} channel-wise subband {DCCRN} with {SNR} estimation for speech enhancement,'' in \emph{Proc. INTERSPEECH}, 2021, pp. 2816--2820.

\bibitem{VCTK+DEMAND}
C.~Veaux, J.~Yamagishi, and S.~King, ``The voice bank corpus: Design, collection and data analysis of a large regional accent speech database,'' in \emph{Proc. Int. Conf. Oriental COCOSDA}, 2013, pp. 1--4.

\bibitem{DEMAND}
J.~Thiemann, N.~to, and E.~Vincent, ``The diverse environments multi-channel acoustic noise database: A database of multichannel environmental noise recordings,'' in \emph{Acoust. Soc. Am.}, vol. 133, no.~5, 2013, p. 3591–3591.

\bibitem{NSNet}
Y.~Xia, S.~Braun, C.~K.~A. Reddy, H.~Dubey, R.~Cutler, and I.~Tashev, ``Weighted speech distortion losses for neural-network-based real-time speech enhancement,'' in \emph{Proc. IEEE Int. Conf. Acoust., Speech, Signal Process.}, 2020, pp. 871--875.

\bibitem{DPCRN}
\BIBentryALTinterwordspacing
X.~Le, H.~Chen, K.~Chen, and J.~Lu, ``{DPCRN:} dual-path convolution recurrent network for single channel speech enhancement,'' \emph{CoRR}, vol. abs/2107.05429, 2021. [Online]. Available: \url{https://arxiv.org/abs/2107.05429}
\BIBentrySTDinterwordspacing

\bibitem{DCRN}
A.~Pandey, C.~Liu, Y.~Wang, and Y.~Saraf, ``Dual application of speech enhancement for automatic speech recognition,'' in \emph{{IEEE} Spoken Language Technology Workshop}, 2021, pp. 223--228.

\bibitem{MTFAA}
G.~Zhang, C.~Wang, L.~Yu, and J.~Wei, ``Multi-scale temporal frequency convolutional network with axial attention for multi-channel speech enhancement,'' in \emph{Proc. IEEE Int. Conf. Acoust., Speech, Signal Process.}, 2022, pp. 9206--9210.

\bibitem{DPTNet}
F.~Dang, H.~Chen, and P.~Zhang, ``{DPT-FSNet}: Dual-path transformer based full-band and sub-band fusion network for speech enhancement,'' in \emph{Proc. IEEE Int. Conf. Acoust., Speech, Signal Process.}, 2022, pp. 6857--6861.

\bibitem{lossSISDR}
J.~L. Roux, S.~Wisdom, H.~Erdogan, and J.~R. Hershey, ``{SDR} - half-baked or well done?'' in \emph{Proc. IEEE Int. Conf. Acoust., Speech, Signal Process.}, 2019, pp. 626--630.

\bibitem{FanSisnr}
T.~Fan, Q.~Li, L.~Shao, Z.~Wu, and A.~Sun, ``Digital phase shift based simulated coherence phase demodulation technology for $\phi$-otdr,'' \emph{Optics Communications}, vol. 546, p. 129746, 2023.

\bibitem{PESQ}
A.~W. Rix, J.~G. Beerends, M.~P. Hollier, and A.~P. Hekstra, ``Perceptual evaluation of speech quality (pesq)-a new method for speech quality assessment of telephone networks and codecs,'' in \emph{Proc. IEEE Int. Conf. Acoust., Speech, Signal Process.}, 2001, pp. 749--752.

\bibitem{STOI}
C.~H. Taal, R.~C. Hendriks, R.~Heusdens, and J.~Jensen, ``A short-time objective intelligibility measure for time frequency weighted noisy speech,'' in \emph{Proc. IEEE Int. Conf. Acoust., Speech, Signal Process.}, 2010, pp. 4214--4217.

\bibitem{Unet}
O.~Ronneberger, P.~Fischer, and T.~Brox, ``U-net: Convolutional networks for biomedical image segmentation,'' in \emph{Medical Image Computing and Computer Assisted Intervention}, vol. 9351, 2015, pp. 234--241.

\end{thebibliography}

\end{document}